\documentclass[referee,sn-basic,Numbered]{sn-jnl}

\usepackage{graphicx}%
\usepackage{multirow}%
\usepackage{amsmath,amssymb,amsfonts}%
\usepackage{amsthm}%
\usepackage{mathrsfs}%
\usepackage[title]{appendix}%
\usepackage{xcolor}%
\usepackage{textcomp}%
\usepackage{manyfoot}%
\usepackage{booktabs}%
\usepackage{algorithm}%
\usepackage{algorithmicx}%
\usepackage{algpseudocode}%
\usepackage{listings}%
\usepackage[normalem]{ulem}

\theoremstyle{thmstyleone}%
\theoremstyle{thmstyletwo}%

\theoremstyle{thmstylethree}%

\usepackage{siunitx}

\newcommand{\Reg}{\mbox{\textit{Re}}_g}

\newcommand{\Fr}{\mbox{\textit{Fr}}_{air}}

\newcommand{\rd}{\sigma}
\newcommand{\f}{\mathtt{F}}
\newcommand{\s}{\mathtt{S}}
\newcommand{\pl}{\mathtt{C}}

\renewcommand{\vec}[1]{\boldsymbol{#1}}

\newcommand{\CR}[1]{\textcolor{black}{#1}}

\newcommand{\ti}[1]{\tilde{#1}}

\newcommand{\tit}[0]{\tilde{t}}

\begin{document}

\title[Bubble curtains in a lock-exchange flow]{Bubble curtains in a lock-exchange flow: the importance of transient dynamics in the curtain-driven regime}


\author[1]{\fnm{Shravan K.R.} \sur{Raaghav}}

\author[1]{\fnm{Ronald J.A.} \sur{Driessen}}

\author[2]{\fnm{Tom S.D.} \sur{O'Mahoney}}

\author[3]{\fnm{Rob E.} \sur{Uittenbogaard}}

\author[1]{\fnm{Herman J.H.} \sur{Clercx}}

\author*[1]{\fnm{Matias} \sur{Duran-Matute}}\email{m.duran.matute@tue.nl}

\affil[1]{\orgdiv{Fluid Dynamics Laboratory and J.M. Burgers Centre, Department of Applied Physics and Science Education}, \orgname{Eindhoven University of Technology}, \orgaddress{\street{PO Box 513}, \postcode{56000MB}, \city{Eindhoven},  \country{The Netherlands}}}

\affil[2]{\orgdiv{Department of Hydraulics for Infrastructure and Industry, Hydraulic Engineering}, \orgname{Deltares}, \orgaddress{\street{Boussinesqweg 1}, \postcode{2629 HV}, \city{Delft}, \country{The Netherlands}}}

\affil[3]{\orgname{Hydro-Key BV}, \orgaddress{\street{Eikendreef 17}, \postcode{6081 EA}, \city{Haelen},  \country{The Netherlands}}}


\abstract{Bubble curtains are line bubble plumes that are used to mitigate saltwater intrusion in ship locks. When the lock gate that separates saline seawater from fresh river water is opened, a lock-exchange flow develops. Installing a bubble curtain at the gate location disrupts this flow and reduces saltwater infiltration. For real-world applications, it is important to quantify how effective the bubble curtain is as a function of the key governing parameters. To this end, we performed multiphase large-eddy simulations that faithfully reproduce earlier experimental results including the two distinct operating regimes: the breakthrough regime and the curtain-driven regime. This paper focuses on the curtain-driven regime and seeks to clarify how the effectiveness of bubble curtains evolves over time. The detailed spatial and temporal data from the simulations, together with the ability to systematically vary the governing parameters, enabled us to overcome several limitations inherent in previous experiments. Furthermore, the simulations were used to obtain parameter values to build a semi-analytical model. Both the simulations and the semi-analytical model successfully capture and elucidate the time evolution of the density field and of the bubble curtain’s effectiveness. The findings highlight that the time elapsed since the gate opening and the  transient dynamics play a crucial role in determining the performance of bubble curtains for mitigation of salt intrusion.}




\maketitle

\section{Introduction}
\label{section:introduction}
Bubble curtains (also known as bubble screens) are line bubble plumes that can be used to mitigate saltwater intrusion in ship/navigation locks \citep{abraham1973pneumatic}. When the lock gate that separates the saline seawater and the fresh river water is opened for ships to pass through, a lock-exchange flow is initiated. This flow is driven by the horizontal density gradient that causes the denser saltwater to flow beneath the lighter fresh water, generating what is known as a \emph{gravity current} or \emph{density current} \citep{benjamin1968gravity, shin2004gravity, huppert2006gravity}. This flow generates \emph{saltwater intrusion} and threatens the availability of fresh water inland from the ship lock. This problem is exacerbated as a consequence of climate change by increasing sea level and more extreme seasonal droughts \citep{Li2025}. The bubble curtain is essentially a bubbly flow that possesses vertical momentum and can stop the horizontal lock-exchange flow, thus mitigating saltwater intrusion. The use of bubble curtains for the mitigation of saltwater intrusion is not limited to locks, but is also used in open systems such as rivers and estuaries \citep{nakai2002experimental, al2021experimental, lu2024experimental,kahrizi2023experimental} where there are opposing flows or tides. 

Bubble curtains have been studied and tested for the mitigation of saltwater intrusion for more than half a century. The first studies by
\citet{abraham1962reduction} and \citet{abraham1973pneumatic} investigated the reduction in salt intrusion using theoretical analysis and field-scale measurement in ship locks with different depths. They proposed the air Froude number $\Fr$ as the main dimensionless parameter governing the flow. This number characterises the strength of the bubble curtain with respect to the strength of the gravity current \CR{(formal definition is provided in Eq. \eqref{eq:non_dimensional_groups})}. In addition, a theoretical solution was proposed to estimate the reduction in saltwater infiltration. This solution agreed reasonably well with the field measurements, but the data points were few and showed large scatter.

Since then, there have been important advances in our understanding, modelling, and measurement techniques of bubbly flows due to their wide range of applications in different engineering disciplines \citep{risso2018agitation, mathai2020bubbly}. They are commonly used in chemical engineering and water treatment applications such as bubble columns, airlift reactors \citep{chisti1987airlift, kantarci2005bubble} and environmental applications other than saltwater intrusion mitigation that include wave breaking \citep{taylor1955action, bulson1961currents}, lake destratification and aeration \citep{wen1987aeration, schladow1993lake, li2025numerical,wang2023effects,murai2025density}, underwater noise mitigation \citep{wursig2000development}, and underwater sediment transport control \citep{cutroneo2014check, wang2024particle,dugue2015influencing,covarrubias2025interaction}, to name a few.

These advances have also allowed us to gain further insight into the workings of bubble curtains. \citet{keetels2011field} performed a combination of laboratory-scale experiments, simulations, and field measurements to test the performance of bubble curtains and compared it with other mitigation measures that included combinations of a bubble curtain with a water jet, a rigged sill, and flushing. \Citet{van2018methods} assessed the design and performance of bubble curtains in a generic lock exchange flow through laboratory experiments with particle image velocimetry to obtain the flow velocity and dye measurements to obtain the density field. The experimental results were then used for validation in the numerical study by \citet{oldeman2020numerical} who described the primary mechanisms behind salt intrusion. They considered six values of $\Fr$ and were the first to accurately capture the optimal value $\Fr\approx 0.91$ that corresponds to the maximum reduction in saltwater intrusion.  \citet{bacot2022bubble} conducted an extensive experimental campaign that covered a wide range of $\Fr$ values. This experimental study is the most detailed to date. They obtained the same optimal $\Fr$ value as \citet{oldeman2020numerical}. They established a formal analogy between air curtains and bubble curtains and defined two regimes of operation of bubble curtains based on the value of $\Fr$. For lower values of $\Fr$, they defined the breakthrough regime in which the inertia of the bubble curtain is not sufficient to counteract the horizontal gravity current flow. As a result, the gravity current flows through the curtain causing a large amount of infiltration. For larger values, the flow was defined as being in the curtain-driven regime, with the bubble curtain forcing two recirculation cells (one on each side of it). In these recirculation cells, the salt water and fresh water are mixed, and secondary gravity currents with intermediate density emerge from them. In addition, \citet{bacot2022bubble} presented an analytical model to estimate the effectiveness of the bubble curtain under steady state conditions in the curtain-driven regime.

All of the studies above \citep{abraham1962reduction,abraham1973pneumatic,keetels2011field,van2018methods,oldeman2020numerical,bacot2022bubble} have mainly considered the performance of bubble curtains under steady-state conditions. However, it is unclear whether or when such conditions are reached, particularly for practical applications in ship locks where the closing of the lock might occur before. There are several time scales that are relevant in the curtain-driven regime, such as the time the bubbles take to rise from the bottom to the surface, the typical time scale for the formation of the recirculation cells and the time that it takes for the secondary gravity currents to reach the end walls of the tank. In fact, \citet{bacot2022bubble} used an approximation to the second one to derive their analytical model. All these time scales depend on different governing parameters of the problem, but their interplay and effect on the effectiveness is unclear before the steady state is reached. Hence, the aim of this study is then to establish a general framework for modelling and understanding the unsteady flow in the curtain-driven regime. We focus on transient dynamics to quantify the effectiveness of the bubble curtain and the reduction of saltwater infiltration at any given time after the opening of the lock gate.

To this end, we investigate the flow in a laboratory-scale lock using a combination of numerical simulations and semi-analytical modelling. In our simulations, we study a wide range of $\Fr$ values for various lock dimensions and a wide range of density differences. With these simulations, we overcome many challenges faced by \citet{bacot2022bubble} in their experiments, such as restrictions on lock length and increased uncertainty in measuring the salt concentration fields in the case of very small density differences. In addition, numerical simulations allowed us to obtain three-dimensional time-resolved velocity and density fields that we use to perform a detailed scaling analysis and determine the necessary parameter values to build a semi-analytical model. This semi-analytical model complements the results from the numerical simulations, providing further insight into the temporal evolution of the flow. In this way, we are able to confirm theoretical derivations by \citet{bacot2022bubble}, but also correct some of their assumptions and expand the theoretical framework.

The remainder of this paper is organised as follows. In Sect. \ref{section:problem_description}, we describe the problem by introducing the relevant non-dimensional parameters and establishing the theoretical preliminaries. The numerical methodology is presented in Sect. \ref{section:numerical_methodology}. This is followed in Sect. \ref{section:curtain_driven_regime} by a discussion of the characteristics of the curtain-driven regime and a scaling analysis using the results of the simulations. In Sect. \ref{section:analytical_modelling}, the semi-analytical model is presented. Then Sect. \ref{section:res_density} and Sect. \S\ref{section:res_effectiveness} present the results of the numerical simulation and the semi-analytical model on the density field and the effectiveness of bubble curtains, respectively. A discussion of the way forward and implications for practical applications is given in Sect. \ref{section:discussion}. Finally, the conclusions are summarised in Sect. \ref{section:conclusions}.

\section{Description of the problem}
\label{section:problem_description}

\begin{figure}
    \centering
    \includegraphics[scale=0.5]{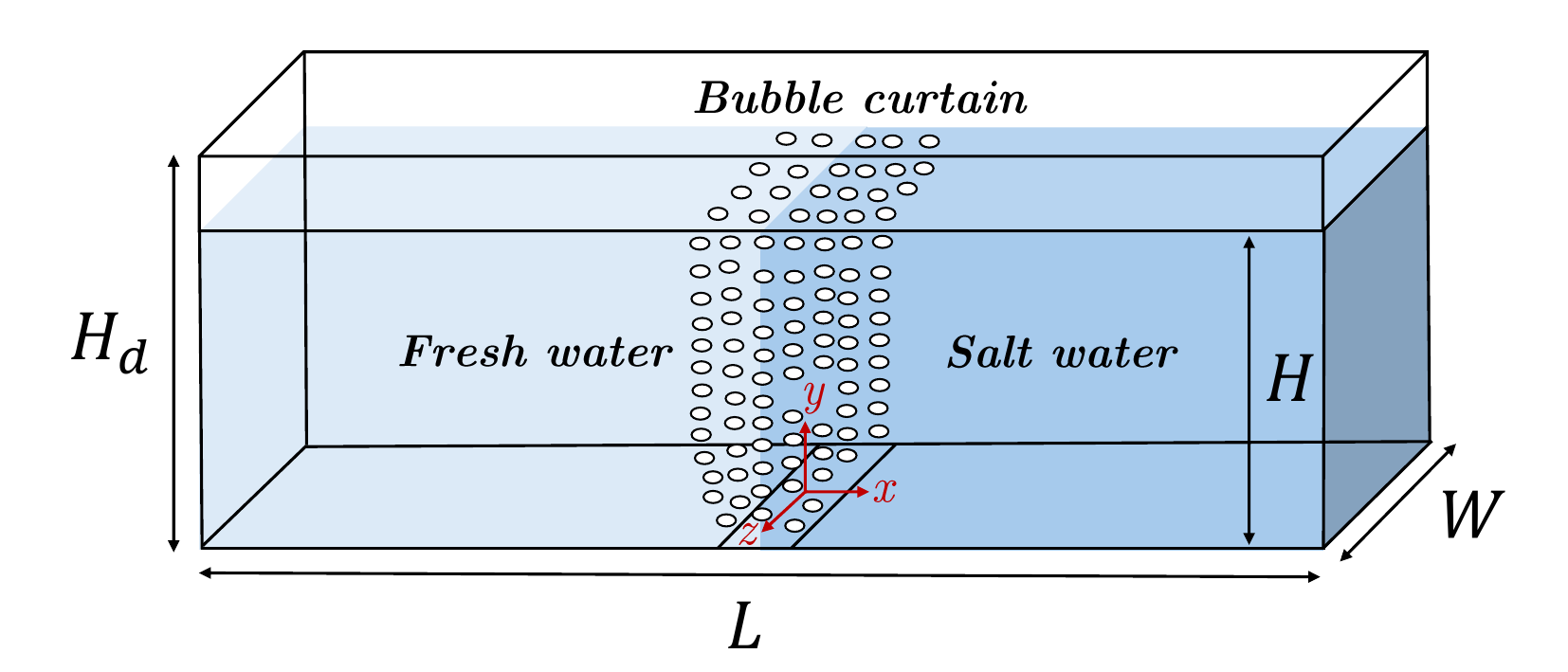}
    \caption{Schematic of a laboratory scale lock with a bubble curtain placed at the centre. The annotations $L$, $W$, and $H_d$ depict the length, width and height of the tank, respectively. The level of the free-surface is $y=H$.}
    \label{fig:general_geometry}
\end{figure}

We consider an idealised laboratory-scale configuration similar to those used by \citet{keetels2011field}, \citet{oldeman2020numerical}, \citet{bacot2022bubble}, and \citet{o2024effect}, as shown in Fig.~\ref{fig:general_geometry}. It consists of a rectangular tank of size $L \times H_{d}  \times W$ in the $x$, $y$, and $z$ directions, respectively. The length $L$ is the largest dimension of the tank. The origin of the reference frame is located at the centre of the tank in the horizontal and at the bottom of the tank in the vertical. The lock gate, located at $x=0$, separates two halves of the tank, each filled with water of different density up to a depth $H$. We define the initial density $\rho_s$ for the denser salt water and $\rho_f$ for the lighter fresh water. Throughout the paper, we refer to the side with initial density $\rho_s$ as the \emph{saltwater side} and the side with initial density $\rho_f$ as the \emph{freshwater side}. The density difference is $\Delta \rho = \rho_s - \rho_f$, and the reduced gravity is $g^{\prime}=g\Delta \rho/\rho_f$ with $g=9.81$ \unit{m.s^{-2}} the gravitational acceleration. The water has a kinematic viscosity $\nu$, which is considered equal for the salt and fresh water.

A porous sparger is located on the tank floor ($y=0$) at the location of the lock gate ($x=0$) and spans the tank in the $z$-direction. In the $x$-direction, it has a width $d_s$, which means that it is located at $- d_s/2 \leq x \leq d_s/2$ and $-W/2 \leq z \leq W/2$. The sparger delivers an air volume flow rate $Q_{air}$ to form the bubble curtain. The bubbles are assumed to have a uniform bubble diameter $d_b$ \footnote{\CR{In the present study, we neglect polydispersity that occurs due to bubble breakup and coalescence. The turbulent Weber number defined as $We_{t}= \rho_{l}\epsilon^{2/3} d_{b}^{5/3}/\gamma$ (where $\epsilon$ is the turbulent dissipation rate \CR{and $\gamma$ the surface tension}) is an important parameter which governs the breakup of bubbles due to the turbulent eddies. Even for the case with highest inlet gas flow rate used in this study, $We_{t}$ was lower than the critical $We_{t}=1.46$ \citep[see][]{martinez1999breakup}. Moreover, it is known from early studies that salt helps stabilize the bubbles and inhibits bubble coalescence \citep{craig1993effect}, thus justifying the uniform (monodisperse) bubble size assumption.}}. We further define the volume air flow rate per unit width $q_{air}=Q_{air}/W$.

Dimensional analysis considering the dimensional variables $L$, $H$, $W$, $\nu$, $q_{air}$, $g$, and $g^{\prime}$, yields five non-dimensional groups:
\begin{equation}
\Reg = \frac{\sqrt{g^{\prime} H}H}{\nu}, \hspace{0.3cm} 
\Fr = \frac{(g \, q_{air})^{1/3}}{\sqrt{g^{\prime} H}}, \hspace{0.3cm}
\rd - 1=\dfrac{g'}{g},\hspace{0.3cm}
\tilde{W} = \frac{W}{H},\hspace{0.3cm}
\tilde{L} = \frac{L}{H},
\label{eq:non_dimensional_groups}
\end{equation}
where $\Reg$ is the Reynolds number of the gravity current, $\Fr$ is the air Froude number introduced in Sect. \ref{section:introduction}, $\rd$ is the relative density of the dense fluid ($\rd=g'/g+1=\rho_s/\rho_f$), and $\tilde{W}$ and $\tilde{L}$ are the aspect ratios of the lock. Although the third dimensionless group is formally $\rd-1$, we chose to work throughout the paper in terms of $\rd$. From the non-dimensional parameters in Eq. \eqref{eq:non_dimensional_groups}, an additional non-dimensional parameter $\hat{q}_{air}=\Fr^3(\rd -1)^{3/2}=q_{air}/(H\sqrt{g H})$ can be defined. This non-dimensional parameter was introduced by \citet{fannelop1991surface} when studying bubble curtains in water with homogeneous density, and they showed that it governs the entrainment characteristics of the curtain. 

Furthermore, we define the dimensionless time $\tilde{t}$ as
\begin{equation}
    \tilde{t}=\frac{t}{\sqrt{H/g'}},
    \label{eq:t_tilde}
\end{equation}
where $t$ is the time elapsed after opening of the lock gate and the typical time scale $\sqrt{H/g'}$ can be interpreted as the time it takes for a saltwater parcel immersed in fresh water to travel from the surface to the bottom \citep{bacot2022bubble}. We will see that this time scale is relevant for several other processes.

For practical purposes, the most common question is to quantify how effective the bubble curtain is in reducing salt intrusion as a function of the relevant non-dimensional parameters. Typically, the effectiveness is presented as a function of $\Fr$ \citep{abraham1962reduction, oldeman2020numerical}. \citet{bacot2022bubble} used instead the deflection modulus $D_{m,b}$ because it is commonly used in the study of air curtains \citep[see e.g.][]{frank2014effectiveness}. However, this is just a preference since these two numbers are related to each other such that
\begin{equation}
D_{m,b} = 2\,\alpha_E\, \Fr^2, 
\label{eq:deflection_modulus}
\end{equation}
where the proportionality constant $\alpha_E$ is the entrainment coefficient of the bubble curtain. \citet{bacot2022bubble} used $\alpha_E=0.071$, which is the entrainment coefficient for turbulent plane plumes according to \citet{paillat2014entrainment}. For convenience, we use only $\Fr$ throughout this paper and not $D_{m,b}$. For all situations considered, $\Reg>12000$ so that the gravity current and flow are always turbulent \citep{simpson1999gravity, shin2004gravity}. Therefore, variations in $\Reg$ are not expected to play an important role, but this parameter is still included here to highlight that the results are valid only for large enough $\Reg$ values, i.e., for turbulent flows. 

In the current article, we chose to keep $\tilde{W}\approx 1.2$ to compare our results with previous studies by \citet{bacot2022bubble}, \citet{oldeman2020numerical}, and \citet{o2024effect}. This value of $\tilde{W}$ is relevant for some ship locks, but others are much wider, with $\tilde{W}$ reaching values close to five, affecting the flow generated by the line bubble plumes \citep{riess1998recirculating}. \CR{The effects of a larger $\tilde{W}$ value as is the case for some real ship locks is discussed in our related study} \citep{raaghav7044620effect}. 

Now, we define the key response parameter, the effectiveness of the bubble curtain, as
\begin{equation}
    E(\tit) = 1 - \frac{V_{bc}(\tit)}{V_o(\tit)},
    \label{eq:effectiveness}
\end{equation}
where $V_{bc}(\tit)$ represents the volume of salt water that has infiltrated at time $\tit$ to the freshwater side of the lock when the bubble curtain is present and $V_o(\tit)$ when the curtain is absent. In other words, $E$ measures the amount of salt water that is blocked by the curtain, $E=1$ representing a perfect barrier and $E=0$ the absence of the curtain. Previous studies \citep{keetels2011field, oldeman2020numerical} have used the salt transmission factor ($STF$) to quantify the effectiveness, but both options give the same information since $STF=V_{bc}/V_o = 1 - E$. \cite{bacot2022bubble} defined the effectiveness in terms of fluxes instead of volumes as is commonly done for air curtains \citep[see e.g.][]{Frank2015}. However, they assumed that these fluxes are constant in time so that the definition of effectiveness becomes identical to Eq. \eqref{eq:effectiveness}. Here, we decided to define the effectiveness directly in terms of total infiltrated volume, since this is commonly the quantity of interest in ship locks. 

The infiltration volume $V_{bc}$ is obtained from mass conservation such that
\begin{equation}
    V_{bc}(\tit) = V_\f\frac{\bar{\rho}_\f(\tit) - \rho_f}{\rho_s - \rho_f},
    \label{eq:v_star}
\end{equation}
where $V_\f=  L \, W \, H/2$ is the volume on the freshwater side of the lock and $\bar{\rho}_\f(\tit)$ is the average density on the same side. Throughout the paper, we will use the symbol $\f$ to denote freshwater side and the symbol $\s$ for the saltwater side. For the infiltrated volume in the absence of a bubble curtain, denoted by $V_o$, we consider a constant flux so that
\begin{equation}
    V_o(\tit) = \dfrac{C_D}{3} W H^2 \tit,
        \label{eq:v_open}
\end{equation}
where $C_D\approx 0.51$ is the discharge coefficient that is obtained by fitting the flow rate obtained in simulations without a bubble curtain (see Appendix \ref{section:Cd_open_cases}). The definition of $V_o$ corresponds to the intrusion of salt water due to the propagation of a gravity current in an infinitely long lock-exchange configuration with a cross-sectional averaged constant velocity $2 C_D \sqrt{g^{\prime}H}/3$ and height $H/2$. Before reaching this velocity, the gravity current undergoes an acceleration phase, but this phase is short compared to the time scales of interest and can be mostly neglected, as will be discussed in Sect. \ref{section:res_density}.

The overarching problem at hand is to predict the characteristics of the flow and, in particular, the effectiveness of the bubble curtain $E$ as a function of the governing parameters $\Fr$, $\rd$, and $\tilde{L}$, and additionally, $\tit$ (while keeping $\tilde{W}$ fixed and $\Reg$ large enough). \citet{bacot2022bubble} has already proposed an analytical model for $E$ as a function of $\Fr$ for $\tit\to\infty$ and $\Fr\gtrsim0.91$. However, most of their experimental results are far from the model prediction. The difference was attributed to the finite length of the tank, which translated to a finite duration of the experiment. To address this problem, we will use numerical simulations and leverage their detailed information to define a semi-analytical model to compute the effectiveness as a function of the three control parameters $(\Fr,\rd, \tilde{L})$ and $\tit$. This methodology allowed us to overcome the main limitations faced by \citet{bacot2022bubble}, while providing detailed time-resolved information.

\section{Numerical methodology}
\label{section:numerical_methodology}
We performed numerical simulations of the setup described in the previous section and shown in Fig.~\ref{fig:general_geometry} for different values of the governing parameters $\Fr$, $\rd$, and $\tilde{L}$ using the open-source finite volume solver OpenFOAM v9 \citep{greenshields2021openfoam}. We used the Euler-Euler two-fluid model to simulate the air-water bubbly upflow \citep{drew1983mathematical, ishii2010thermo} and the air-water interface at $y\approx H$. In this approach, the air and water phases are assumed to be interpenetrating continua, and the momentum exchange between the phases is modelled using a closure force term in the momentum equations. We employ large-eddy simulations to account for the turbulence generated as a result of the bubbly flow and the gravity current.  

\subsection{Governing equations and boundary conditions}
\label{section:governing_equations}
We solve the governing phase-averaged continuity equation and Navier-Stokes equations in the Boussinesq approximation for incompressible Newtonian fluids \citep{drew1983mathematical}:
\begin{equation}
    \frac{\partial\alpha_{q}}{\partial t} +{\boldsymbol{\nabla}}\cdot (\alpha_q \vec{v}_q) = 0,
    \label{eq:bouss_continuity}
\end{equation}
\begin{equation}
\begin{split}
    \frac{\partial(\alpha_q \vec{v}_q)}{\partial t} +\boldsymbol{\nabla}\cdot\left(\alpha_q\vec{v}_q \vec{v}_q \right) = -\frac{\alpha_q}{\rho_{0,q}}{\boldsymbol{\nabla}} p + \frac{\rho_q}{\rho_{0,q}} \alpha_q \vec{g} \\ + {\boldsymbol{\nabla}} \cdot \left\{\alpha_q \nu_{e,q}\left[{\boldsymbol{\nabla}}\vec{v}_q + ({\boldsymbol{\nabla}}\vec{v}_q)^T -\frac{2}{3}({\boldsymbol{\nabla}}\cdot \vec{v}_q){\boldsymbol{I}}\right] \right\} + \vec{F}_q.
    \label{eq:bouss_momentum}
    \end{split}
\end{equation}
where the subscript $q$ denotes the phase [$g$ for gas (air) or $l$ for liquid (water)], $\alpha_q$ represents the volume fraction for each phase, $\vec{v}_q=(v_{q,x},v_{q,y},v_{q,z})$ the associated flow velocity, $\rho_q$ the density, $\rho_{0,q}$ a reference density, $\nu_{e,q}$ the effective kinematic viscosity (i.e. the sum of the molecular viscosity $\nu_q$ and the eddy viscosity $\nu_{t,q}$), and $\vec{F}_q$ the momentum transfer between the phases. Furthermore, $p$ is the pressure that is shared between the two phases and $\vec{g}= (0,-g,0)$ is the gravitational acceleration. The momentum transfer term $\vec{F}_q$ is an action-reaction pair that is used to model the interaction between the phases. To model the bubble curtain, we consider the drag, lift, and virtual mass forces on the bubbles and the corresponding reaction on the liquid phase. These forces are modelled as briefly explained in the description of our solver in Sect. \ref{section:solver}. Since we are mainly interested in the liquid phase and to simplify the notation, we obviate the subindex $l$ for this phase so that $\nu_l=\nu$, $\rho_l=\rho$, $\alpha_l=\alpha$, $\vec{v}_l=\vec{v}$, and so on. 

To account for the density difference as a result of variations in the salt concentration, the Boussinesq approximation was used in Eq. \eqref{eq:bouss_momentum} for the liquid phase. According to this approximation, variations in density are only incorporated into the gravitational term in Eq. \eqref{eq:bouss_momentum} \citep{boussinesq1903theorie}. Commonly, the hydrostatic balance is subtracted from the vertical component of Eq. \eqref{eq:bouss_momentum} so that only a small density perturbation is multiplying the gravitational acceleration \citep[see e.g.][]{nieuwstadt2016introduction}. However, this is not done here because of the particular way equations are implemented in the solver. The Boussinesq approximation is only valid for flows with small density variations \citep{nieuwstadt2016introduction}, which is a fair assumption in this study, since $\Delta \rho/\rho_f\leq 0.04$. \CR{We define $s$ (with $0\leq s \leq 1$) as the intrinsic salt water fraction with density $\rho_s$ and the corresponding salt water fraction constrained to the liquid phase as
\begin{equation}
    \alpha_s = \alpha s.
\end{equation}
The density of the liquid is then
\begin{equation}
    \rho = \rho_f + s(\rho_s-\rho_f) = s \rho_s + (1 - s) \rho_f,
\end{equation}
where we have considered the density of fresh water as the reference density ($\rho_{0}=\rho_f$) such that the perturbation density is $\rho' = \rho - \rho_f=s(\rho_s-\rho_f)$.} Because the diffusivity of salt in the gas phase is orders of magnitude lower than in the liquid phase and the salt does not cross the interface between the two phases, the transport of salt is assumed to be present only in the liquid phase. In the gas phase, the density is then constant, so $\rho_g=\rho_{0,g}$, and therefore buoyancy effects within this phase are not present.

Since the salt is constrained to the liquid phase, its transport is obtained by solving the scalar transport equation for the field $\alpha_s$,
\begin{equation}
    \frac{\partial(\alpha s)}{\partial t} +\boldsymbol{\nabla}\cdot\left(\alpha\vec{v} s\right)-
    \boldsymbol{\nabla} \cdot \CR{\left(\alpha D_{eff} \boldsymbol{\nabla} s\right)} = 0,
    \label{eq:salt_transport}
\end{equation}
where $D_{\textit{eff}} = D + D_{t}$ is the effective diffusivity with $D$ the molecular diffusivity and $D_t$ the turbulent or eddy diffusivity. The turbulent diffusivity $D_t$ is related to the eddy viscosity such that
\begin{equation}
    D_t = \nu_{t} / Sc_t,
\end{equation}
where $Sc_t$ is the turbulent Schmidt number. A constant value $Sc_t=0.7$ was used following, for example, \citet{spalding1971concentration} and \citet{oldeman2020numerical}. \citet{spalding1971concentration} demonstrated that this value yielded very good agreement between experiments and simulations when predicting the concentration fluctuation profiles in a round turbulent free jet. For the current flow configuration, \citet{oldeman2020numerical} also observed very good agreement with the experiments using this value, thereby justifying this choice.

The values of the kinematic viscosity of water $\nu=1.123\times 10^{-6}$ \unit{m^2.s^{-1}} (variations in $\nu$ due to differences in salt concentration are neglected), surface tension $\gamma = 0.0734$ \unit{N.m^{-1}} and salt diffusivity $D=1.096\times 10^{-9}$ \unit{m^2.s^{-1}} were estimated based on an ambient temperature of 288.5 \unit{K}, which corresponds to the density of fresh water $\rho_f = 999$ \unit{kg.m^{-3}}. For the gas phase (i.e., the air in the bubbles and on top of the water surface), we considered the same ambient temperature that yields a density $\rho_{g}=1.223$ \unit{kg.m^{-3}} and a kinematic viscosity $\nu_g=1.464\times 10^{-5}$ \unit{m^2.s^{-1}}. Assuming a constant air density in the bubbles is a good approximation in small-scale experiments, but compressibility effects might have to be considered to model field-scale bubble curtains due to the large depth.

The domain has three types of boundaries, as seen in Fig.~\ref{fig:general_geometry}. These are the top boundary at $y = H_{d}$, the side walls at $x = -L/2$, $x = L/2$, $z = -W/2$, and $z = W/2$, and the bottom boundary at $y=0$. This last boundary is a wall boundary, except in the location of the sparger. At these boundaries, we need to specify boundary conditions for the velocity, the gas/liquid volume fraction, and the salt water fraction. For the latter, we apply, all around the domain, a no-flux condition: $ \vec{\hat{n}} \cdot \nabla s =0$ with $\vec{\hat{n}}$ the unit vector normal to the boundary. For the bottom and side walls, no-slip and no-penetration boundary conditions are applied for the velocity ($\vec{v}_q=0$), and a zero-gradient boundary condition is imposed on the volume fractions of the phases ($\vec{\hat{n}} \cdot \nabla \alpha_q  =0$). At the sparger, air enters the tank to form the bubble curtain, and hence a velocity-inlet condition is specified. The inlet velocity of water is set to 0 \unit{m.s^{-1}}, while the inlet velocity of air in the $y$-direction is $v_{g,y} = Q_{air}/(d_s\, W)$ with $d_s=0.02$ \unit{m}, while setting $\alpha_g$ = 1.

Finally, the top boundary at $y=H_d$ represents the air above the setup (e.g., above the tank in an experimental setup). Here, the velocity gradient normal to the boundary is set to zero $(\partial \vec{v}_q/\partial y=0$) when the flow is directed out of the domain. However, if the velocity is directed into the domain, the velocity is estimated and enforced to maintain mass balance. In addition, the condition for the volume fractions is a zero-gradient condition if the flow is directed out of the domain $(\partial \alpha_q/\partial y = 0)$, but when the flow is directed into the domain, the values $\alpha_g=1$ and $\alpha=0$ are enforced. This boundary condition allows air to leave and enter the domain, while water can leave but cannot enter the domain. However, water never leaves through this boundary because the air layer separating it from the water surface is sufficiently thick.

\subsection{Numerical solver}
\label{section:solver}
The governing equations \eqref{eq:bouss_continuity}, \eqref{eq:bouss_momentum}, and \eqref{eq:salt_transport} are discretised and solved using large eddy simulations (LES). Due to the open-source nature of OpenFOAM and the code written in a highly object-orientated fashion, the solver can be customised to suit specific applications without modifying basic libraries or classes \citep{weller1998tensorial, jasak2007openfoam, chen2014openfoam}. The default Euler-Euler multiphase solver included in the compilation is the multiphaseEulerFoam solver \citep{rusche2003, wardle2013hybrid}. In the present study, this solver was customised to include the capabilities to simulate bubbly flows in the presence of active scalars such as salt. We added the phase-constrained transport equation for the salt transport and implemented the Boussinesq approximation. The governing equations presented in Sect. \ref{section:governing_equations} are strongly coupled and therefore stiff, which complicates the solution procedure and demands robust numerical methods along with a careful choice of discretization schemes. This is necessary to maintain solution stability and strict boundedness of the scalar fields. Below, we describe the specific configuration used.

OpenFOAM offers three options for temporal schemes, namely Euler, Crank-Nicolson, and backward time stepping schemes. The Euler scheme is by nature a bounded time-stepping scheme but is first-order accurate. Because we wanted a second-order accurate solver in both space and time, we adopted the Crank–Nicolson scheme \citep{crank1947practical}. Nevertheless, when using this scheme, the \CR{salt water fraction} $s$ frequently became unbounded, particularly for higher inlet air flow rates. Hence, we also used a flux correction technique, which is an OpenFOAM's implementation of Zalesak's iterative flux-corrected transport algorithm \citep{zalesak1979fully}, to solve the salt transport equation Eq. \eqref{eq:salt_transport}. \CR{Here, we use the sequential operator-splitting procedure \citep{hundsdorfer2003numerical} where the purely advective transport of Eq. \eqref{eq:salt_transport}, which is prone to spurious over- and under-shoots, is first solved using the explicit MULES technique to ensure boundedness. The resulting salinity field is used as the initial field, to march the remaining implicit diffusive transport in Eq. \eqref{eq:salt_transport}. A dynamic time stepping approach was employed, where the time step was computed based on the maximum Courant number, which was set to 0.5.}

For spatial discretization, second-order TVD (Total Variation Diminishing) schemes were used to ensure stability even at high air flow rates \citep{van1974towards, greenshieldsweller2022}. The solver is pressure based, and therefore the pressure-correction method was employed to solve Eqs. \eqref{eq:bouss_continuity} and \eqref{eq:bouss_momentum}. We used the PIMPLE algorithm \citep{greenshieldsweller2022} which is a combination of the widely used PISO and SIMPLE algorithms \citep{ferziger2002computational}. For the outer or predictor loop, where the transport equations of the volume fraction are solved and the momentum equations are optionally predicted, the number of correctors was set to three. For the inner loop, where the pressure Poisson equation is solved and the momentum equation is corrected, the number of correctors was also set to three. \CR{The pressure equation was solved with a geometric agglomerated algebraic multigrid (GAMG) solver, and the equations for the velocities, the salt water fraction and the SGS kinetic energy with a symmetric Gauss--Seidel solver, with a solver tolerance of $10^{-9}$ for all fields.}

To model the effects of the interaction between the gas and liquid phases on the turbulence, we use two subgrid-scale models. For the gas phase, the continuousGasKEqn model was used, which is a one-equation subgrid-scale model. This is an unpublished model developed for Eulerian multiphase solvers and available in OpenFOAM. For the liquid phase, the one-equation model proposed by \citet{niceno2008one} was used. 

When using LES together with Eulerian two-fluid modelling, bubble size and grid size are coupled by the Milelli criterion \citep{milelli2002numerical}, which states that the mesh size should be at least 1.5 times the diameter of the bubbles. In physical terms, for the model to represent the dispersed bubbles, they should occupy less than one grid cell since the equations are phase averaged and the dispersed phase is treated as a continuum. The bubble diameter, $d_b= 2$ mm, was chosen in agreement with the bubble size in previous laboratory experiments \citep{bacot2022bubble, o2024effect}. Hence, a structured grid of 5 \unit{mm} was used, except for the case with the largest lock dimensions, where a 10 \unit{mm} grid was used due to computational constraints. For the drag and lift forces, we used the closures proposed by \citet{schiller1933drag} and \citet{tomiyama2002transverse}, respectively. In line with \citet{rzehak2017unified} and \citet{oldeman2020numerical}, we used a constant virtual mass coefficient of 0.5. This choice is further supported by the results of \citet{zhang2006numerical} who showed that in bubble columns there is a negligible difference when either a constant virtual mass coefficient or one that depends on the bubble aspect ratio are used. \CR{Note that at the air-water free surface neither phase is the dispersed phase and hence the interphase closures are handled with blending factors across the two phases that depend on the local phase fraction fields \citep{rusche2003, wardle2013hybrid}. The air–water free surface is represented implicitly by the phase volume-fraction fields with a finite-thickness transition in the phase volume fraction. The interface thickness is hence dependent on the spatial resolution and numerical schemes.}

 To test the correct implementation of our solver, we compared the results of our solver with previous results. Specifically, to test the correct implementation of the Boussinesq approximation, we compared with the results from the classical lock exchange simulations by \citet{hartel2000}; to test the correct implementation of the bubbly flow, we compared with the bubble column simulations by \citet{deen2001large} and \citet{deen2001experimental} and the fresh-fresh case by \citet{oldeman2020numerical}. Our simulations successfully reproduced the flow behaviour observed in these test cases. Finally and most importantly, we compare the complete implementation throughout the paper with the experimental results of \citet{bacot2022bubble} and the numerical results of \citet{oldeman2020numerical}. Although it is not feasible to exactly replicate every experimental detail (such as the sparger design or the manual gate removal), the aim is to reproduce the main flow features and dynamics, and, as will be demonstrated, our solver achieves this. \CR{To avoid a very lengthy discussion here, more details on the grid size requirements and other numerical treatments related to the solver are left for a follow up article focussed on numerical modelling.}

\subsection{Parameter values and initialization procedure}
\label{section:flow_initialization}

We performed simulations covering a wide range of $\Fr$, $\tilde{L}$, and $\rd$ values: $0.34\leq \Fr\leq 3.79$, $7.7\leq \tilde{L}\leq 40.0$, and $1.003\leq \rd \leq 1.040$. This range of parameters is large enough to include typical values that are found in real ship locks and previous experimental work. In this paper, we will focus on a subset with $1.07\leq \Fr\leq 3.79$, $10.0\leq \tilde{L}\leq 40.0$, and $1.003\leq \rd \leq 1.038$. To achieve these parameter values, we varied $L$, $\Delta\rho$, $H$, and $q_{air}$. We consider three values for $L$ (2, 4, and 6 m) to overcome some of the finite-size effects experienced by \citet{bacot2022bubble} in their 2m-long tank. To vary $\Delta \rho$, only the saltwater density $\rho_s$ was varied while the freshwater density was kept constant at $\rho_f = 999$ \unit{kg.m^{-3}}. We employed a range of values for $H$ between 0.15 m and 0.7 m. To keep $\tilde{W}\approx 1.2$  (formally $1.125\leq \tilde W \leq 1.33$), $W$ was varied with $H$. \CR{Note that since $\tilde{W}$ varies within such a small range, this minimal change does not affect the flow characteristics \citep{raaghav7044620effect}.} The variation of $\Delta \rho$ and $H$ implies that $\Reg$ was also varied since $\Reg \propto \Delta \rho^{1/2}H^{3/2}$. However, since the gravity current and flow are fully turbulent with $\Reg>12 000$ \citep{simpson1999gravity, shin2004gravity}, variations in $\Reg$ hardly affect our results as will be seen later.

For all simulations performed, the dimensionless volume flow rate $\hat{q}_{air}$ spans the range $2.8\times10^{-2}\gtrsim \hat{q}_{air} \gtrsim 1.1\times10^{-4}$. To compare the simulations with the experimental data and characterise the regimes presented in Sect. \S\ref{section:characteristics}, the entire range $\hat{q}_{air}$ is considered. However, for the scaling analysis presented in Sect. \S\ref{section:scaling_analysis} and semi-analytical modelling in Sect. \S\ref{section:analytical_modelling}, we discarded the simulations with $\hat{q}_{air}<0.004$ because we observed that the entrainment characteristics of the bubble curtain started to change around that value, with the results of those simulations deviating from the general trends \citep{raaghav2026breakthrough}. \CR{Note that the limit of $\hat{q}_{air} \approx 0.004$ adopted here is a slightly conservative value compared to the one reported by \citet{raaghav2026breakthrough}, where the dependence of the entrainment characteristics was observed for $\hat{q}_{air} \lesssim 0.002$.}

It is important to note that there are limits on the values of $q_{air}$, $\Delta\rho$ and $H$ in the simulations. The maximum value of $q_{air}$ was 0.005 \unit{m^2 s^{-1}}. Beyond this value, the effects of polydispersity due to bubble breakup and coalescence can become significant and cannot be neglected \citep{kantarci2005bubble}. As discussed in Sect. \ref{section:governing_equations}, the maximum value of $\rho_s$ used is limited due to the Boussinesq approximation. The grid size poses a restriction on the minimum $H$ used due to the resolution requirements to model gravity currents using LES \citep{pelmard2018grid} and the Milelli criterion to solve for the bubbly flow. Hence, a minimum $H$ of 0.15 \unit{m} was used.

Each simulation was performed in two steps, using a procedure similar to that of \citet{oldeman2020numerical}. We first simulate the bubble curtain in a lock filled on both sides with fresh water for 30 \unit{s}. During this time, the bubble curtain develops, reaches the water surface, and establishes a quasi-steady recirculating flow on both sides. At $t=30$ \unit{s}, salt water is patched on the saltwater side of the lock by instantaneously making $s = 1$, setting the density equal to $\rho_s$ there. Afterwards, the simulation is resumed. \CR{This procedure mimics the operation of bubble curtains in laboratory experiments \citep{bacot2022bubble, o2024effect} and in practice, where the curtain is switched on before the lock gate is opened.}
For domain lengths 2, 4, and 6 \unit{m}, the total simulation time (including the initialisation of 30 \unit{s}) was chosen a priori to be 60 \unit{s}, 80 \unit{s} and 100 \unit{s}, respectively. The end times were chosen to allow the secondary gravity currents to reach the end walls without extending the simulations much further. However, because the exact time for gravity currents to reach the end wall was \emph{a priori} unknown, the simulation time was slightly shorter in some cases.

\section{The curtain-driven regime: characteristics and scaling analysis}
\label{section:curtain_driven_regime}

\subsection{Characteristics}
\label{section:characteristics}

\begin{figure}
    \centering
    \includegraphics[width=1\textwidth]{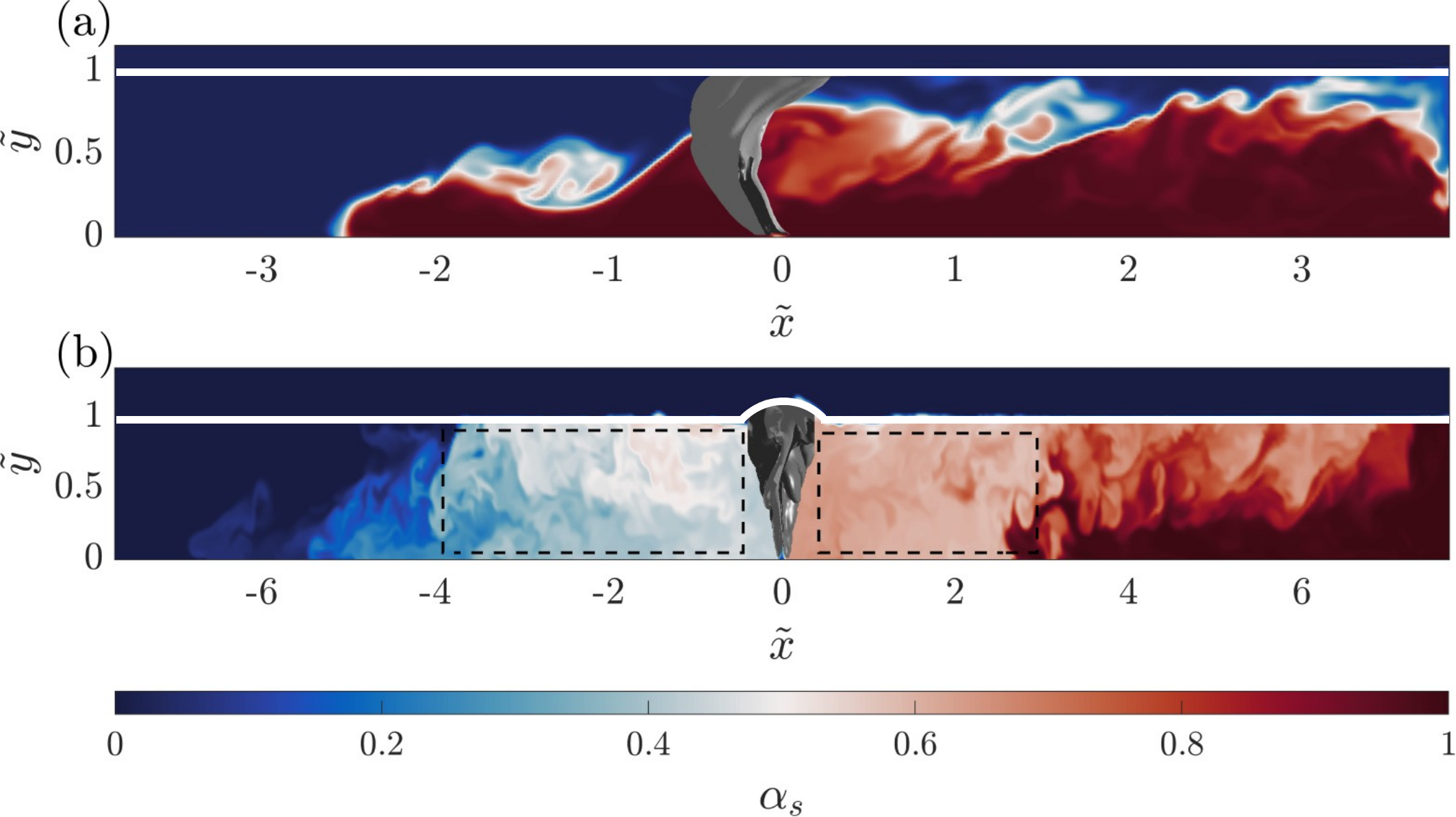}
    \caption{Snapshots of contours of the phase-constrained salt water fraction $\alpha_s$ at $z = 0$ for two simulations: a) one in the breakthrough regime ($\Fr=0.5$, $\rd=1.040$, $\tilde{L}=7.69$) and b) one in the curtain-driven regime ($\Fr=3.24$, $\rd=1.005$, $\tilde{L}=15.4$). The location of the bubble curtain is shown with grey shading. \CR{Here, $\tilde{x} = x/H$ and $\tilde{y} = y/H$.} For simulation in the curtain-driven regime, we highlight the two recirculation cells with intermediate density using rectangles. On the freshwater side, the cell extends until $\tilde{x}\approx-4$ and on the saltwater side until $\tilde{x}\approx3$. From each of these recirculation cells, a secondary gravity current emerges.}
    \label{fig:regimes}
\end{figure}

For this flow, two distinct regimes, the breakthrough and curtain-driven regimes, have been identified depending on the value of $\Fr$ \citep{bacot2022bubble}. Examples of these two regimes are shown using fields of the salt water fraction $\alpha_s$ in Fig.~\ref{fig:regimes}. In the breakthrough regime ($\Fr\lesssim 0.91$), the inertia of the gravity current dominates over that of the bubble curtain. Hence, the gravity current breaks through the bubble curtain and salt water intrudes through the curtain (see Fig.~\ref{fig:regimes}\emph{a} and supplementary movie 1). In the curtain-driven regime ($\Fr\gtrsim 0.91$), the inertia of the bubble curtain dominates that of the gravity current and therefore acts as a barrier between the saltwater and freshwater sides (see Fig.~\ref{fig:regimes}\emph{b} and supplementary movie 2). This second regime is the focus of this paper. In it, the bubble curtain continuously entrains fluid from both sides, creating a recirculation cell on each side (i.e., one for $x<0$ and one for $x>0$). From these cells, secondary gravity currents emerge and propagate towards the end walls of the tank. 

\begin{figure}
    \centering
    \includegraphics[width=0.98\textwidth]{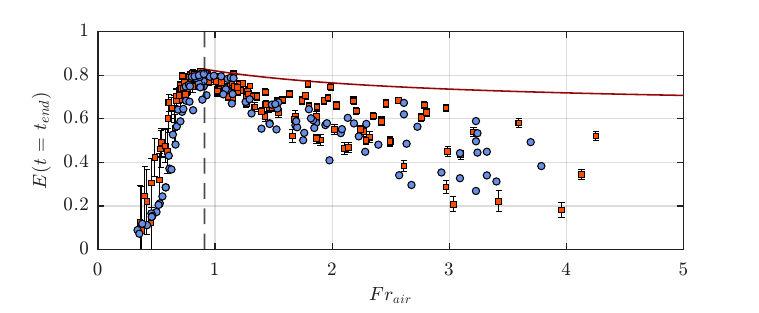}
    \caption{Plot of effectiveness $E$ vs Froude number $\Fr$, comparing the simulation results (blue circles) obtained in the present study and experimental data of \citet{bacot2022bubble} (red squares). The effectiveness is computed at the end time $t_{end}$ of the experiments or simulations. The analytical solution proposed by \citet{bacot2022bubble} for $t_{end}\to \infty$ is shown by the solid red curve. The dashed line represents the optimal $\Fr$ ($\approx 0.91$) where $E$ is maximum that separates the breakthrough and curtain-driven regimes which are typically observed to the left and right of the dashed line, respectively.} 
    \label{fig:E_validation}
\end{figure}

\citet{bacot2022bubble} computed the effectiveness at the time when the gravity current reached the wall of the tank ($t=t_{end}=t_w$) for each experiment as shown in Fig.~\ref{fig:E_validation}. In addition, this figure shows the effectiveness as obtained from our simulations at $t=t_{end}$. For the simulations, $t_{end}=t_w$ if the secondary gravity current reaches the end wall of the tank within the simulation time. Otherwise, $t_{end}$ is the end time of the simulation. The experiments and simulations show similar trends and accurately capture the optimal $\Fr$ value ($\Fr\approx 0.91$) where $E$ reaches a maximum ($E\approx0.81$ for both cases). This optimum marks the transition between the breakthrough and curtain-driven regimes.

In general, the flow characteristics of the two observed regimes are consistent with those reported by \citet{bacot2022bubble} and the optimal value $\Fr\approx 0.91$ is recovered in the simulations (Fig.~\ref{fig:E_validation}). In addition, both experiments and simulations show similar trends in the values of $E$ as a function of $\Fr$ in the breakthrough regime ($\Fr\lesssim 0.91$), and the simulation results fall within the error bars of the experimental results, giving us confidence in that the relevant processes are well reproduced by the simulations. The seemingly small differences between experiments and simulations in the breakthrough regime \CR{are discussed in our related study \citep{raaghav2026breakthrough}}. Here, instead, we focus on the curtain-driven regime ($\Fr\gtrsim 0.91$), where there is an apparently random scatter in the values of $E(\tit=\tit _{end})$. This scatter prevents us from observing a clear one-to-one correspondence between simulations and experiments. \CR{For this reason, we compare the effectiveness values for a couple of cases where we exactly matched all the problem parameters with \citet{bacot2022bubble} such that all the non-dimensional groups in Eq. \eqref{eq:non_dimensional_groups} have the same values. The results, presented in Table \ref{table:effectiveness_comparison} in Appendix \ref{section:appendix_validation}, show an excellent agreement with the experimental data.} In addition, we compared the time evolution of the average density in the recirculation cells as obtained from simulation and experiments showing very good agreement (see Appendix \ref{section:appendix_validation} for two examples). This agreement gives us confidence on the accuracy of our simulations in reproducing physical processes in the curtain-driven regime.

Clearly, the scatter in $E(\tit=\tit_{end})$ within the curtain-driven regime presents a challenge in linking the effectiveness to the value of $\Fr$. Furthermore, we note that the experimental and numerical results are mostly far from the analytical model proposed by \citet{bacot2022bubble} for $t_{end}\to \infty$. They already observed the scatter and hypothesised that it is due to the finite domain size of the lock, which determines in part the value of $\tit_{end}$. This hypothesis indicates that $E$ is strongly dependent on time. Here, we aim to support this hypothesis, explain the scatter, and quantify the dependence of the effectiveness on time  using our simulations and a relatively simple semi-analytical model.

\subsection{Scaling analysis}
\label{section:scaling_analysis}
\begin{figure}
    \centering
    \includegraphics[scale = 0.18]{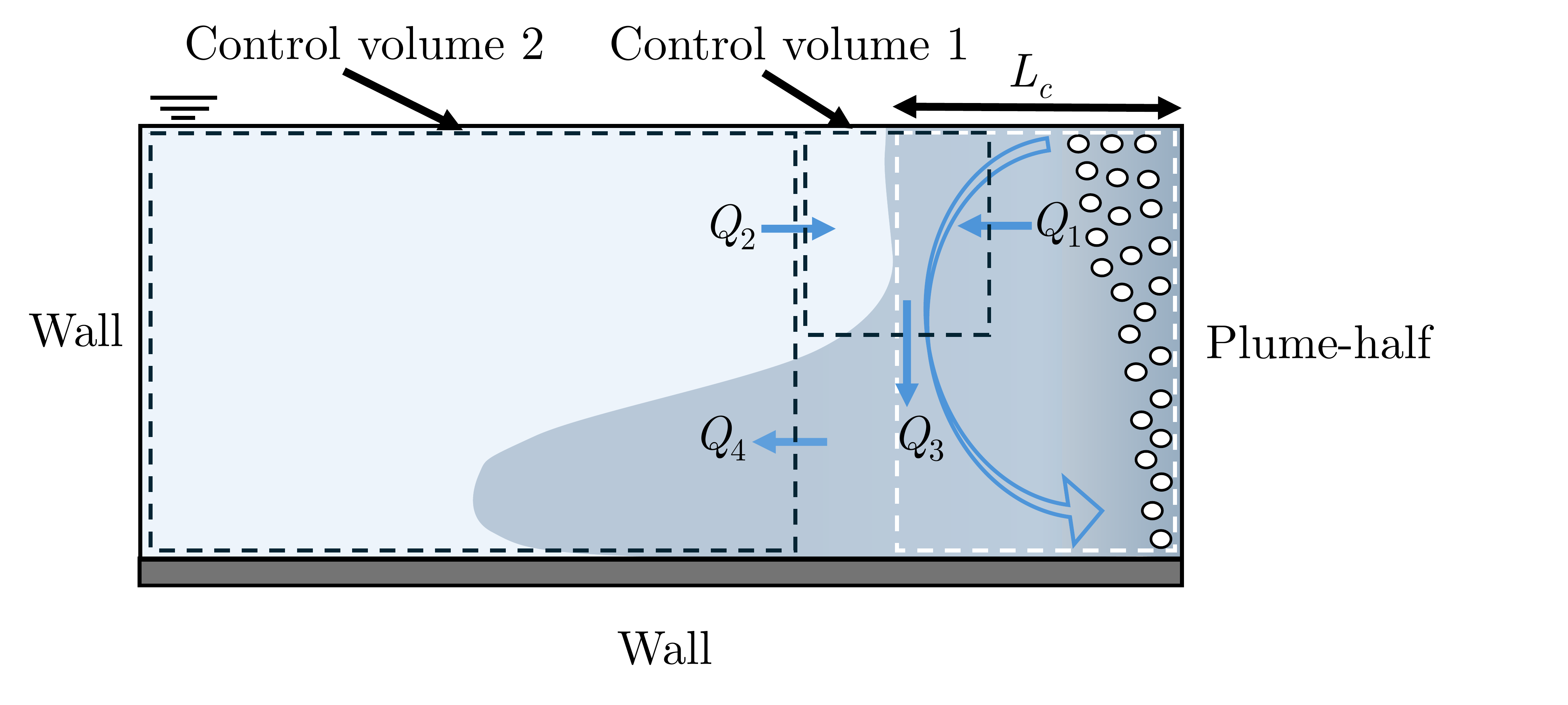}
    \caption{Schematic representation of the recirculation cell (shown using dashed white box) with length $L_c$ and the secondary gravity current on the freshwater side. The control volumes chosen to establish mass and volume conservation are also depicted.} 
    \label{fig:schematic_recirculation_cell}
\end{figure}

As a first step, we perform a scaling analysis to derive the typical velocity, time, and length scales of the recirculation cells  expanding and \CR{refining} the approach by \citet{bacot2022bubble}. As will be shown, these typical time scales will help us interpret our results. An intuitive schematic of the flow field for the freshwater side of the lock is shown in Fig.~\ref{fig:schematic_recirculation_cell}. For the sake of brevity, only this side of the lock is shown, but a similar analysis can be done on the saltwater side. For our analysis, we consider two control volumes and the volume flow rates through their faces, which are denoted by $Q_j$ with $j=1\dots4$. Furthermore, we consider that these flow rates have reached a steady state (i.e. they are constant). An in-depth discussion on the actual time needed to reach such constant flow rates and the correctness of our assumption will follow below.

As discussed in Sect. \ref{section:problem_description}, the typical velocity scales of a line bubble plume and the gravity current are $(gq_{air})^{1/3}$ and $\sqrt{g^{\prime}H}$, respectively. Therefore, it is logical to expect that a typical velocity scale $U_c$ in the recirculation cell should contain contributions of both of these scales. To obtain the functional relation between $U_c$, on the one hand, and $(gq_{air})^{1/3}$ and $\sqrt{g^{\prime}H}$, on the other, we assume two control volumes as depicted in Fig.~\ref{fig:schematic_recirculation_cell}. First, we consider the control volume 1 centred around the boundary of the recirculation cell and the fresh water. Each face of control volume 1 has the same area $A$, and they extend throughout the width $W$ of the tank.  From the diagram, we assume that $U_c\sim Q_3/A$ and that
\begin{equation}
 Q_3=Q_{1} + Q_{2} ,
    \label{eq:mass_conservation_velocity_scale}
\end{equation}
where $Q_1$ and $Q_2$ are the volume flow rates through the opposite lateral faces of control volume 1.

To estimate these flow rates, we assume that the velocity of the surface current generated by the curtain scales with its typical velocity, that is, $Q_1/A \propto (gq_{air})^{1/3}$. To estimate $Q_2$, we use mass conservation in control volume 2. Due to the small variation in density, we consider this volume to be constant (i.e., neglecting surface deformations) such that $Q_{2} = Q_{4}$. In other words, the volume flow rate due to the secondary gravity current emanating from the recirculation cell is the same as the volume flow rate of fresh water into the recirculation cell in the upper half of the domain. Since the recirculation cell has an intermediate mixed density ($\bar{\rho}_c$), the density difference driving the gravity current is $\bar{\rho}_c - \rho_f$. Since $\bar{\rho}_c$ is a mixture of water with density $\rho_s$ and $\rho_f$, we assume here that this mixture is of equal parts such that $\bar{\rho}_c - \rho_f\approx \Delta \rho/2$. The values for $\bar{\rho}_c - \rho_f$ obtained from the simulations and from the semi-analytical model are presented later in Sect. \S\ref{section:res_density}, so the accuracy of this estimate can be checked a posteriori. For the scaling analysis at hand, we consider it suitable yielding
\begin{equation}
Q_{2}/A = Q_{4}/A \propto \sqrt{g^{\prime}H},
    \label{eq:volume_conservation_velocity_scale}
\end{equation}
and 
\begin{equation}
    \label{eq:velocity_scale}
    U_c  \sim \dfrac{Q_3}{A} =  C_1 (gq_{air})^{1/3}+C_2\sqrt{g^{'}H},
\end{equation}
where $C_1$ and $C_2$ are proportionality constants. The velocity scale in dimensionless form is then given by
\begin{equation}
\tilde{U}_c\equiv\frac{U_c}{\sqrt{g^{'}H}} \sim C_1 \Fr+C_2.
\label{eq:velocity_cell_dimensionless}
\end{equation}

\begin{figure}
    \centering  \includegraphics[width=0.49\textwidth]{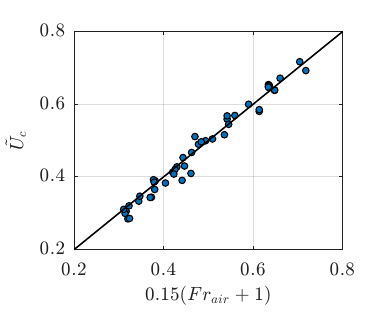}
 \caption{Non-dimensional velocity scale in the recirculation cell, $\tilde{U}_c$, computed using Eq. \eqref{eq:velocity_scale_flux} versus the proposed scaling in Eq. \eqref{eq:velocity_cell_dimensionless_2}. The solid black line represents a perfect match with $\tilde{U}_c= 0.15( \Fr+1)$.} 
    \label{fig:velocity_scale_comparison}
\end{figure}

We verify whether the scaling given by Eq. \eqref{eq:velocity_cell_dimensionless} is observed in the simulations. We define the velocity scale from the simulations as
\begin{equation}
U_c = \frac{1}{L_c W}\displaystyle \int_{z=-W/2}^{z=W/2} \int_{x=-L_c}^{x=0}|u_y(x,H/2,z)| dxdz,
\label{eq:velocity_scale_flux}
\end{equation}
where $L_c$ is the length of the recirculation cell and the velocity $u_y$ has been averaged over time considering only times after a steady state has been established in the recirculation cell. The values of $C_1$ and $C_2$ are obtained using multiple linear regression yielding $\CR{0.157 \pm 0.007}$ and $\CR{0.13\pm0.02}$ (with 95\% CI), respectively. Henceforth, for simplicity, we consider the approximation\CR{, which is supported by a one-parameter fit of the form $\tilde{U}_c = a(\Fr+1)$ yielding $a=0.149\pm0.002$:}
\begin{equation}
\tilde{U}_c\approx 0.15( \Fr+1).
\label{eq:velocity_cell_dimensionless_2}
\end{equation}
Figure \ref{fig:velocity_scale_comparison} presents the left side of this equation versus the right side for all of our simulations, showing good agreement.

The next step is to determine a typical time scale. Since the parcels are continuously lifted up and down in the recirculation cell over the water height with typical velocity $U_c$, we define the typical mixing time scale inside the cell $\tau_{mix}\equiv H/U_c$ such that in dimensionless form
\begin{equation}
\tilde{\tau}_{mix}\equiv \dfrac{\tau_{mix}}{\sqrt{H/g'}} \approx \frac{1}{0.15(\Fr+1)}.
\label{eq:time_scale}
\end{equation}
This is the second time scale originally mentioned in the introduction and should give an indication of the time taken for the water in the recirculation cells to become homogeneous, and for the flow in the recirculation cells to become steady. A further interpretation of the physical meaning of $\ti{\tau}_{mix}$ and a discussion of its importance is given in Sect. \ref{section:res_density}.

 \begin{figure}
    \centering
    \includegraphics[width= 0.98\textwidth]{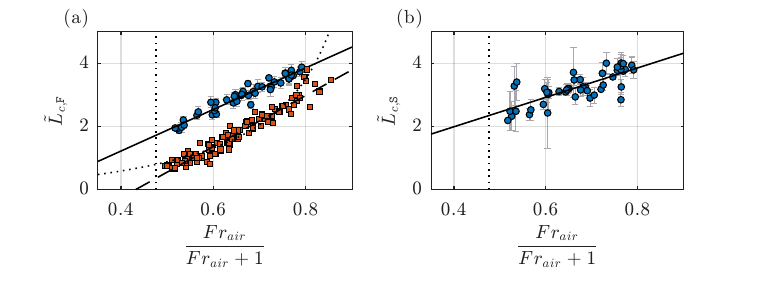}
    \caption{Dimensionless length of the recirculation cell $\tilde{L}_c$ as a function of $\Fr/(\Fr+1)$ for the simulations (blue circles) and for the experiments by \citet{bacot2022bubble} (red squares). The error bars represent $\pm 1$ standard deviation and the vertical dotted line the $\Fr$ value for optimal effectiveness ($\Fr=0.91$) for reference. The black solid lines are linear fits for the simulation results given by Eqs. \eqref{eq:scaling_Lc_left} and \eqref{eq:scaling_Lc_right}. The dashed line is the linear fit to the experiments by \citet{bacot2022bubble}: $\tilde{L}_{c,\f}=(8.1\pm0.4)\Fr/(\Fr+1)-(3.5\pm0.3)$. The dotted curve represents the scaling proposed by \citet{bacot2022bubble} to their data: $\tilde{L}_c \approx 0.62\sqrt{2}\Fr$. 
    } 
\label{fig:L_cell_non_dimensional}
\end{figure}

Now that we have established the typical velocity and time scales, they can be used to compute the length scale of the recirculation cell $L_c$ (see Fig.~\ref{fig:schematic_recirculation_cell}). In dimensional form, $L_c$ is proportional to the product of the surface current velocity produced by the bubble curtain (with this velocity proportional to $(gq_{air})^{1/3}$) and $\tau_{mix}$ such that $L_c \propto (gq_{air})^{1/3}\tau_{mix} $. In non-dimensional form,
\begin{equation}
\tilde{L}_c\equiv\frac{L_c}{H} \propto \Fr\ti{\tau}_{mix}\propto \frac{\Fr}{\Fr + 1}.
\label{eq:length_scale_non_dimensional}
\end{equation}
Figure \ref{fig:L_cell_non_dimensional} shows the length of the recirculation cell $\tilde{L}_c$ obtained from the simulations on both sides of the lock as a function of $\Fr/(\Fr+1)$ (see Appendix \ref{section:appendix_L_cell} for the detailed procedure to compute $L_c$). 
For the freshwater side (Fig.~\ref{fig:L_cell_non_dimensional}a), the fit yielded
\begin{equation}\label{eq:scaling_Lc_left}
    \tilde{L}_{c,\f}=
    (6.6\pm 0.5)\dfrac{\Fr}{\Fr+1}-(1.4\pm 0.3).
\end{equation}
and for the saltwater side (Fig.~\ref{fig:L_cell_non_dimensional}\emph{b}), it yielded
\begin{equation}\label{eq:scaling_Lc_right}
    \tilde{L}_{c,\s}= (4.65\pm 1.1)\dfrac{\Fr}{\Fr+1}+(0.1\pm 0.8),
\end{equation}
where the error represents the \CR{95\% confidence interval (CI)}.

From Eqs.~\eqref{eq:scaling_Lc_left} and \eqref{eq:scaling_Lc_right}, it can be seen that the recirculation cells have approximately the same size (certainly within the error margins given by the 95\% CI), but a difference with $\ti{L}_{c,\f}\lesssim \ti{L}_{c,\s}$ starts to become apparent for $\Fr\CR{\lesssim 2}$\CR{, with the difference growing as $\Fr$ decreases}. In fact, on the freshwater side, we can see from Eq. \eqref{eq:scaling_Lc_left} that $\ti{L}_{c,\f}=0$ occurs for a finite value $\Fr\approx \CR{0.27}$. The fact that the recirculation cell vanishes for a finite value of $\Fr$ is consistent with the existence of the breakthrough regime, where this recirculation cell is absent. However, the value of $\Fr$ at which $\ti{L}_{c,\f}=0$ should be considered with care, since the fit does not consider data for $\Fr<1.07$, and a different trend might occur then. In the opposite limit, when $\Fr \rightarrow \infty$, the situation is equivalent to the case where the lock contains only fresh water. In that limit, \CR{$L_{c,\f} \approx 5.2H$ and $L_{c,\s} \approx 4.8H$} (i.e. $L_c\propto H$) according to Eqs. \eqref{eq:scaling_Lc_left} and\eqref{eq:scaling_Lc_right}, respectively, which is in good agreement with previous studies on bubble curtains in water with homogeneous density \citep{fannelop1991surface, riess1998recirculating}. For the saltwater side of the tank, the scatter around the fit is larger than for the freshwater side because the limit of the recirculation cell is not as clearly defined. However, the data still show good agreement with the scaling given by Eq. \eqref{eq:length_scale_non_dimensional} and the recirculation cell is approximately the same size as the one on the freshwater side. 

Figure \ref{fig:L_cell_non_dimensional}(\emph{a}) also shows the experimental results for $\tilde{L}_{c,\f}$ by \citet{bacot2022bubble}. The values of $\ti{L}_{c,\f}$ from our simulations are higher than those of the experiments by \citet{bacot2022bubble}. This kind of quantitative differences between the experiments and the simulations could be expected, since certain characteristics of the experiments are not faithfully reproduced by the simulations. We suspect that one key difference is the sparger, which is a pipe with equally spaced holes in the experiments but represents porous material in the simulations. The second key difference might be the bubble size.  \citet{bacot2022bubble} did not measure the bubble size in the experiments, but estimated it between 2 mm and 7 mm. The bubbles modelled in the simulations are in the lower limit of this range and hence are probably smaller than in the experiments. These two differences are important because they affect the entrainment by the bubble curtain, which directly affects the recirculation cell length \citep{fannelop1991surface, o2024effect}.

Even if the experimental values for $\tilde{L}_{c,\f}$ are smaller than for the simulations, they also collapse onto a line when plotted as a function of $\Fr/(\Fr+1)$ with a slope ($8.1\pm0.4$) similar to that obtained for our simulations. In their scaling analysis, \citet{bacot2022bubble} considered that the velocity in the recirculation cell was due only to buoyancy of the water parcels, while ignoring the effect of the bubble curtain. In that case, they obtained $\tilde{L}_{c,\f}\propto \Fr$, and a linear fit using data with $L_{c,\f}<0.5$ m yielded $\tilde{L}_{c,\f}\approx0.62 \sqrt{2} \Fr$ (shown as a dotted curve in Fig.~\ref{fig:L_cell_non_dimensional}\emph{a}). They observed that this fit shows satisfactory results for small $L_{c,\f}$ and $\Fr$ values, but overestimated $L_{c,\f}$ for larger $\Fr$ values. This is observed in Fig.~\ref{fig:L_cell_non_dimensional}(\emph{a}) as the steep increase in the dotted curve for $\Fr/(\Fr+1)\gtrsim 0.75$ (i.e. $\Fr \gtrsim 3$). Hence, the fact that they did not consider the effect of the bubble curtain means that their scaling underestimates $\tau_{mix}$ and overestimates $L_{c,\f}$ for large $\Fr$ values, while our scaling analysis provides an appropriate description across the entire range of $\Fr$ values.

\section{Semi-analytical model}
\label{section:analytical_modelling}
\begin{figure}
    \centering
    \includegraphics[width=\linewidth]{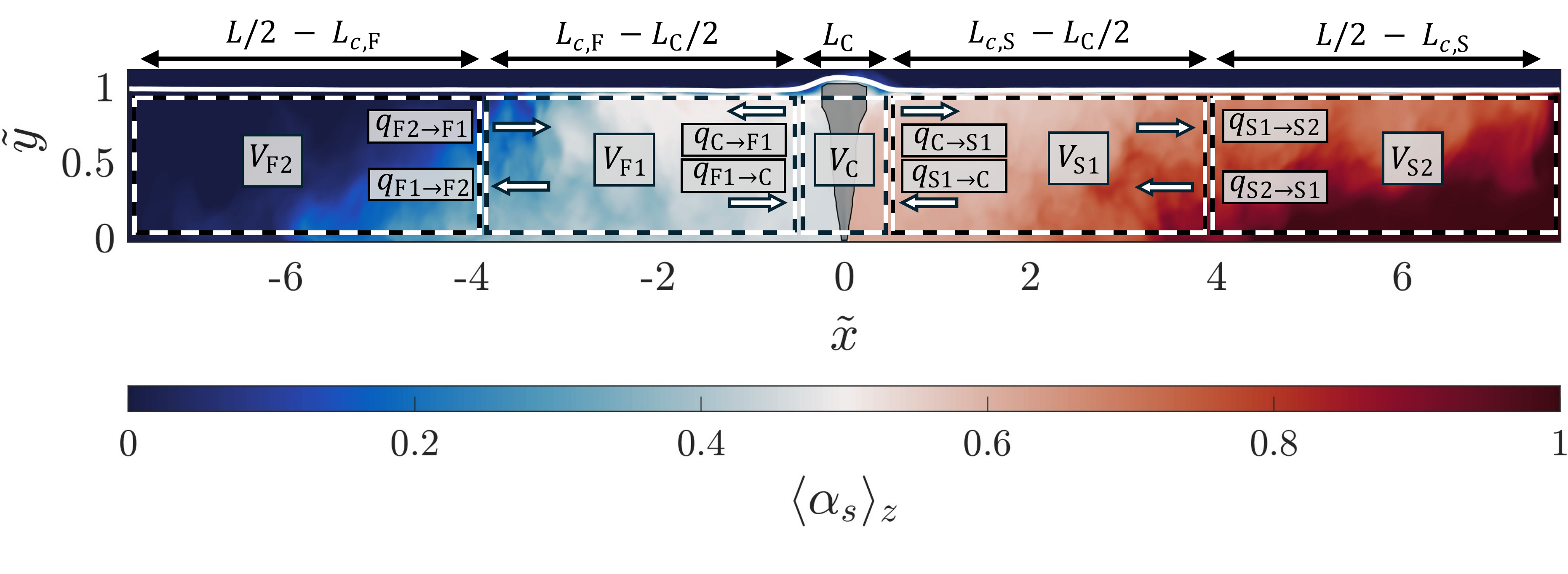}
    \caption{Schematic diagram of the control volumes chosen to derive the semi-analytical model. The control volumes are marked with rectangles using dashed lines and their labels annotated. The length of the control volumes is shown at the top. The labels corresponding to the width-averaged volume flow rates through the faces are shown. \CR{Here, $\tilde{x} = x/H$ and $\tilde{y} = y/H$.} The colour in the background represents the width-averaged salt water fraction $\langle \alpha_s \rangle_z$ for the simulation also shown in Fig.~\ref{fig:regimes}b, and the gray shading represents the location of the bubble curtain.}
    \label{fig:analytical_model_sketch}
\end{figure}

In this section, we introduce the semi-analytical model to estimate the effectiveness in the curtain-driven regime as a function of time. We build on the steady-state derivation by \citet{bacot2022bubble}, but avoid certain assumptions by making use of the data from the simulations, thus further extending it. Furthermore, since this study focusses on the transient dynamics, the model is not limited to the steady state but is applicable at any given time. This model is not intended as a predictive model for the effectiveness for any lock and bubble curtain configuration, since the parameter values are, in principle, only valid for the particular configuration of the simulations. Instead, it should be seen as a surrogate model that allows us to overcome limitations (e.g. a finite domain) and avoid performing more and costlier numerical simulations and laboratory experiments (e.g. in larger domains and for longer times).

To describe the model, we use the salt water fraction field for a simulation with $\Fr = 3.24$ also shown in Fig.~\ref{fig:analytical_model_sketch}.  We consider five control volumes. Starting from the centre, the control volume that encompasses the bubble curtain is indicated by the subindex $\pl$ and extends from $x=-0.35H$ to $x=0.35H$, so $V_\pl=L_\pl H W$ with $L_\pl=0.7H$. The exact length of this control volume is not critical for the model, but it should encompass most of the curtain without extending far into the recirculation cells on each side. On each side of the central control volume, we define the control volumes that encompass the recirculation cells. These are indicated as $V_{\f 1}$ for the freshwater side and $V_{\s 1}$ for the saltwater side. They extend to the edges of the recirculation cells at $x=-L_{c,\f}$ on the left and $x=L_{c,\s}$ on the right, as given by Eqs. \eqref{eq:scaling_Lc_left} and\eqref{eq:scaling_Lc_right}, respectively. Hence, $V_{{\f 1}}=(L_{c,\f}-L_\pl/2)HW$ and $V_{{\s 1}}=(L_{c,\s}-L_\pl/2)HW$. Finally, at the extremes of the tank, the remaining volumes are $V_{\f 2}=(L/2 - L_{c,\f})HW$ for the freshwater side and $V_{\s 2}=(L/2 - L_{c,\s})HW$ for the saltwater side. We neglect the variation in $H$ due to the free surface deformation so that the control volumes are constant.

 For each control volume, we define the spatially averaged density inside it as $\bar{\rho}_j(t)$ and the volume flow rate per unit width between two control volumes as $q_{j\to i}$, where the subindices $j$ and $i$ refer to the control volumes $\f 1$, $\f 2$, $\pl$, $\s 1$ or $\s 2$. Since each control volume and the total volume are constant, $|q_{j\to i}|= |q_{i\to j}|$. For flow rates, we assume the sign convention that positive flow rates enter the control volume of interest and negative flow rates exit it.
 
In the three central control volumes ($\f1, \pl, \s 1$), we assume that the water is well mixed so that the density is homogeneous and $\rho=\bar{\rho}_j(t)$ with $j$ representing the control volume $\f1, \pl,$ or $\s 1$. Hence, at $x=- L_\pl/2$, there is a flow rate $q_{\f 1\to \pl}$ of water with density $\bar{\rho}_{\f1}$ close to the bottom \citep[for $y/H\lesssim 3/4$ as also found by][]{abraham1973pneumatic} and a flow rate $q_{\pl \to \f 1}$ of water with density $\bar{\rho}_\pl$ close to the surface. Similarly, at $x= L_\pl/2$, there is a flow rate $q_{\s 1 \to \pl}$ of water with density $\bar{\rho}_{\s1}$ close to the bottom (also for $y/H\lesssim 3/4$) and a flow rate $q_{\pl \to \s 1}$ of water with density $\bar{\rho}_\pl$ close to the surface. For the control volume $\f 2$ at the left end, the secondary gravity current emanates from $\f 1$ with density $\bar{\rho}_{\f 1}$ and flow rate $q_{\f1\to \f2}$ in the bottom half, and due to mass conservation there is a flow rate $q_{\f 2 \to \f1}$ of water with density $\rho_f$ close to the surface. For the control volume $\s 2$ at the right end, the secondary gravity current emanates from $\s 1$ close to the surface with density $\bar{\rho}_{\s 1}$ and flow rate $q_{\s1\to \s2}$, and there is a flow rate $q_{\s 2 \to \s1}$ of water with density $\rho_s$ close to the bottom. Hence, on the saltwater side of the lock, the surface current generated by the curtain and the secondary gravity current are aligned in the top half of the lock. However, on the freshwater side, the secondary gravity current is generated at the bottom, and by volume conservation (see Sect. \ref{section:scaling_analysis}), this creates a flow against the surface current generated by the curtain on the top part of the lock. Hence, there is an asymmetry between the freshwater and saltwater sides of the lock. 

The above description for the three central volumes can be expressed in non-dimensional form using mass conservation as the following system of ordinary non-linear differential equations:
\begin{align}
\tilde{L}_{\pl} \dfrac{d\tilde{\bar{\rho}}_{\pl}}{d\tilde{t}}& =  -|\tilde{q}_{\f 1 \to \pl}|(\tilde{\bar{\rho}}_{\pl}-\tilde{\bar{\rho}}_{\f 1})- |\tilde{q}_{\s 1 \to \pl}|(\tilde{\bar{\rho}}_{\pl}-\tilde{\bar{\rho}}_{\s 1}),\label{eq:model_a} \\
\tilde{L}_{\f1}\dfrac{d\tilde{\bar{\rho}}_{\f 1}}{d\tilde{t}} & = |\tilde{q}_{\f 1 \to \pl}|(\tilde{\bar{\rho}}_{\pl}-\tilde{\bar{\rho}}_{\f 1})+ |\tilde{q}_{\f 1 \to \f 2}|(1-\tilde{\bar{\rho}}_{\f 1}),\label{eq:model_b}\\
\tilde{L}_{\s1} \dfrac{d\tilde{\bar{\rho}}_{\s 1}}{d\tilde{t}} & =  |\tilde{q}_{\s 1 \to \pl}|(\tilde{\bar{\rho}}_\pl-\tilde{\bar{\rho}}_{\s 1})+ |\tilde{q}_{\s 1 \to \s 2}|( \rd-\tilde{\bar{\rho}}_{\s 1}),\label{eq:model_c}
\end{align}
where the tildes denote non-dimensional variables with $\tilde{L}=L/H$, $\tilde{\rho} = \rho/\rho_f$, $\tilde{t} = t/\sqrt{H/g^{\prime}}$, $\tilde{q} = q/\sqrt{g^{\prime}H^3}$, and $\rd=\rho_s/\rho_f$. The length of the control volumes are computed by subtracting $\tilde{L}_\pl/2$ to the length of the recirculation cells given by Eqs. \eqref{eq:scaling_Lc_left} and \eqref{eq:scaling_Lc_right}. In this way, we define
\begin{equation}
    \tilde{L}_{\f1}  = \tilde{L}_{c,\f}-\tilde{L}_\pl/2=   
    6.6\dfrac{\Fr}{\Fr+1} - 1.75
\label{eq:length_cv_left}
\end{equation}
and
\begin{equation}
    \tilde{L}_{\s1}  = \tilde{L}_{c,\s}-\tilde{L}_\pl/2=4.65\dfrac{\Fr}{\Fr+1}-0.25.
\label{eq:length_cv_right}
\end{equation}
Notice that $\tilde{L}_\pl/2$ is smaller than the errors in the estimations of the length of the recirculation cells. Hence, small changes in the definition of the $\pl$ control volume should not affect the overall accuracy of the semi-analytical model. 

\begin{figure}
    \centering
\includegraphics[width=0.9\textwidth]{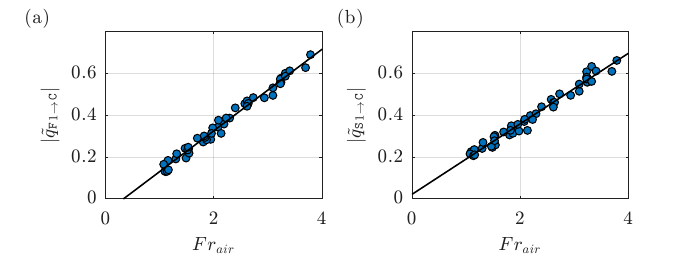}
\caption{Volume flow rate per unit width a) $|\tilde{q}_{\f 1 \to \pl}|$ and b) $|\tilde{q}_{\s 1 \to \pl }|$ as a function of $\Fr$. Data from simulations is shown with blue bullets and the black lines represent the linear fits Eqs. \eqref{eq:fluxes_fit_1} and \eqref{eq:fluxes_fit_2}.}
\label{fig:fluxes1}
\end{figure}

To solve for the densities in the three central control volumes ($\f 1$, $\pl$ and $\s 1$) using Eqs. \eqref{eq:model_a}--\eqref{eq:model_c}, we still need estimates of the volume flow rates $\tilde{q}_{\f 1\to \pl}$, $\tilde{q}_{\s 1\to \pl}$, $\tilde{q}_{\f 1\to \f 2}$, and $\tilde{q}_{\s 1\to \s 2}$. For this estimation, we used the results of the numerical simulations. The volume flow rates in and out of the plume control volume $q_{\f1 \to \pl}$ and $q_{\s 1\to \pl}$ are dominated by entrainment due to the bubble curtain and should scale with the typical magnitude of the associated velocity; thus $ (gq_{air})^{1/3}H$ \citep{bulson1961currents}. Hence, $\tilde{q}_{\f1 \to \pl} \sim \tilde{q}_{\s 1\to \pl} \propto \Fr$. In Fig.~\ref{fig:fluxes1}, we plot $\tilde{q}_{\f1 \to \pl}$ and $\tilde{q}_{\s 1 \to \pl}$ from simulations against $\Fr$. The approach to compute the volume flow rate amongst control volumes from the simulations is described in Appendix \ref{section:appendix_flux_compute}. We observe linear relationships between the flow rates and $\Fr$. 

Linear fits to the data are also shown in Fig.~\ref{fig:fluxes1} corresponding to \CR{
\begin{equation}
|\tilde{q}_{\f 1 \to \pl}| = |\tilde{q}_{\pl \to \f 1}| = (0.19 \pm 0.01 ) \Fr -(0.06\pm 0.02)
\label{eq:fluxes_fit_1}
\end{equation}}
for the freshwater side, and 
\begin{equation}
    |\tilde{q}_{\pl \to \s 1}| = |\tilde{q}_{\s 1 \to \pl}| = (0.17 \pm 0.01) \Fr + (0.02\pm 0.02),
    \label{eq:fluxes_fit_2}
\end{equation}
for the saltwater side. Using the results of \citet{bulson1961currents} and assuming symmetry such that $\tilde{q}_{\pl \to \s 1}=\tilde{q}_{\f 1 \to \pl}$, \cite{bacot2022bubble} estimated $\tilde q\approx 0.18 \Fr$, which is close to our findings, except for small values of $\Fr$.   

\begin{figure}
    \centering
\includegraphics[width=0.9\textwidth]{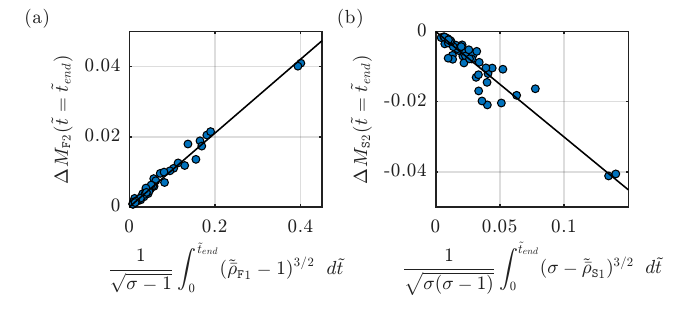}
    \caption{Change in mass in control volumes a) $\f 2$ and b) $\s 2$ plotted against the respective estimates from Eqs. \eqref{eq:fluxes_fit_4a} and \eqref{eq:fluxes_fit_4b} for all the simulations. The solid black lines represent the linear fits Eqs. \eqref{eq:fluxes_fit_4a} and \eqref{eq:fluxes_fit_4b}.} 
    \label{fig:fluxes2}
\end{figure}

The volume flow rates due to the secondary gravity currents are proportional to their propagation speed. Notice that the secondary gravity current in the freshwater side has density $\bar{\rho}_{\f1}$ and propagates into an ambient fluid with density ${\rho}_f$ so that its propagation speed is proportional to $\sqrt{g(\bar{\rho}_{\f 1}-\rho_f)H/\rho_f}$. For the saltwater side, the secondary gravity current has density $\bar{\rho}_{\s 1}$ and propagates into an ambient fluid with density $\rho_s$ so that its propagation speed is proportional to $\sqrt{g(\rho_s-\bar{\rho}_{\s 1})H/\rho_s}$. By dividing the densities by $\rho_f$ and using $H$ as the typical length scale, the volume flow rates are then given in dimensionless form by
\begin{align}
|\tilde{q}_{\f 1 \to \f 2}| & = |\tilde{q}_{\f2 \to \f 1}| = \alpha_{gc}\sqrt{\frac{\tilde{\bar{\rho}}_{\f 1}-1}{\rd-1}},\label{eq:fluxes_fit_3a}\\
|\tilde{q}_{\s1 \to \s2}| & = |\tilde{q}_{\s2 \to \s 1}| = \beta_{gc}\sqrt{\frac{\rd- \tilde{\bar{\rho}}_{\s 1}}{\rd(\rd-1)}},
\label{eq:fluxes_fit_3b}
\end{align}
with $\alpha_{gc}$ and $\beta_{gc}$ positive constants. \CR{Although $\tilde{\bar{\rho}}_{\f 1}$ and $\tilde{\bar{\rho}}_{\s 1}$ depend on time, secondary gravity currents only emerge after a certain time when the salinity field in the recirculation cells is homogenised. Hence we approximate $\alpha_{gc}$ and $\beta_{gc}$ as constants.} To obtain an estimate for $\alpha_{gc}$ and $\beta_{gc}$, we consider the amount of changes in mass $\Delta M_{\f 2}$ and $\Delta M_{\s 2}$ in the control volumes $\f 2$ and $\s 2$ that are related to the flow rates following
\begin{align}
\Delta M_{\f 2}(\tilde{t}) & =  \int_{0}^{\tilde{t}} (\tilde{\bar{\rho}}_{\f 1}(\tilde{t}') - 1) \tilde{q}_{\f 1 \to \f 2}(\tilde{t}') d\tilde{t}' = \alpha_{gc}\int_{0}^{\tilde{t}} (\tilde{\bar{\rho}}_{\f 1}(\tilde{t}') - 1) \sqrt{\frac{\tilde{\bar{\rho}}_{\f 1}(\tilde{t}') - 1}{\rd-1}}d\tilde{t}' \nonumber \\ & =  \frac{\alpha_{gc}}{\sqrt{\rd-1}} \int_{0}^{\tilde{t}} (\tilde{\bar{\rho}}_{\f 1}(\tilde{t}') - 1)^{3/2} d\tilde{t}',\label{eq:fluxes_fit_4a}\\
\Delta M_{\s 2}(\tilde{t}) & = \int_{0}^{\tilde{t}} (\tilde{\bar{\rho}}_{\s 1}(\tilde{t}') - \rd) \tilde{q}_{\s 1 \to \s 2}(\tilde{t}') d\tilde{t}' = \CR{\beta_{gc}}\int_{0}^{\tilde{t}} (\tilde{\bar{\rho}}_{\s 1}(\tilde{t}') - \rd) \sqrt{\frac{\rd -\tilde{\bar{\rho}}_{\s 1}(\tilde{t}')}{\rd(\rd-1)}}d\tilde{t}' \nonumber \\ & =   -\frac{\beta_{gc}}{\sqrt{\rd(\rd-1)}} \int_{0}^{\tilde{t}} (\rd -\tilde{\bar{\rho}}_{\s 1}(\tilde{t}'))^{3/2} d\tilde{t}'\label{eq:fluxes_fit_4b},
\end{align}
with $\tilde{t}'$ a dummy variable. Figure \ref{fig:fluxes2} shows the left side  vs. the right side of Eqs. \eqref{eq:fluxes_fit_4a} and \eqref{eq:fluxes_fit_4b} (omitting the constant factors $\alpha_{gc}$ and \CR{$-\beta_{gc}$}) at $\tilde{t}=\tilde{t}_{end}$ as obtained from the simulations. In other words, it shows the changes in mass in the control volumes $\f 2$ and $\s 2$ as a function of the mass that has been transported to those control volumes. A linear relationship is observed, and the fits yield $\alpha_{gc}=0.105\pm0.003$ and $\beta_{gc}= \CR{0.30\pm0.02}$. The differences in the coefficients obtained between the freshwater and saltwater sides are the result of the asymmetry between both sides, which is then incorporated into our model. On the saltwater side, the gravity current and surface current aid each other, resulting in $\beta_{gc}>\alpha_{gc}$. We further note the larger scatter and error on the saltwater side, which results from the fact that the recirculation cell is less clearly defined on this side.  Here, it is convenient to note that $\rd(\rd-1) \approx (\rd-1)$ since $\rd \approx 1$ for all our simulations to satisfy the requirements of the Boussinesq approximation. If this simplification is used, all density differences are normalised by $\rd - 1$. However, we keep the full expression in our model.

Having established relations for the volume flow rates and length of the control volumes in Eqs. \eqref{eq:length_cv_left}--\eqref{eq:fluxes_fit_3b}, we can now solve the system of equations Eqs. \eqref{eq:model_a}--\eqref{eq:model_c}, and it can be immediately seen that the solution depends on both $\rd$ and $\Fr$. In the following section, a detailed description of this dependence will be provided.  The system of non-linear ordinary differential equations is solved numerically with initial conditions $\tilde{\bar{\rho}}_{\pl}(\tilde{t}=0) = (1 + \rd)/2$, $\tilde{\bar{\rho}}_{\f 1}(\tilde{t}=0) = 1$ and $\tilde{\bar{\rho}}_{\s 1}(\tilde{t}=0)=\rd $. 

From the averaged densities as a function of time obtained from solving Eqs. \eqref{eq:model_a}--\eqref{eq:model_c}, we compute the effectiveness. Recalling the definitions Eqs. \eqref{eq:effectiveness}--\eqref{eq:v_open}, the effectiveness can be written in non-dimensional form as
\begin{equation}
    E(\tilde{t}) = 1 - STF =1 - \frac{V_{bc}(\tilde{t})}{V_{o}(\tilde{t})} = 1 - \frac{3\tilde{L}(\tilde{\bar{\rho}}_{\f} - 1)}{2C_D(\rd-1)\tilde{t}},
    \label{eq:effectiveness_analytical}
\end{equation}
with the average density in the freshwater side given by
\begin{equation}
\tilde{\bar{\rho}}_{\f}(\tilde{t}) = \dfrac{2}{\tilde{L}}(\Delta M_{\f 1} + \Delta M_{\f 2} + \Delta M_{\pl}/2+\tilde{L}/2),
\end{equation}
where $\Delta M_i$ represents the amount of mass gained in the control volume $i$,
so that 
\begin{equation}
    \Delta M_{\f 1}(\tit) = (\tilde{\bar{\rho}}_{\f 1}- 1)\tilde{L}_{\f 1},
    \label{eq:DeltaML1}
\end{equation}
\begin{equation}
    \Delta M_{\pl}(\tit) = [\tilde{\bar{\rho}}_{\pl}- (\rd+1)/2]\tilde{L}_{\pl},
    \label{eq:DeltaMP}
\end{equation}
and $\Delta M_{\f 2}$ is given by Eq. \eqref{eq:fluxes_fit_4a}. \CR{Note that $\Delta M_\pl$ is measured with respect to the initial density $(\rd+1)/2$ assumed in the model for the control volume $\pl$, and not with respect to $\rho_f$. In this way, $V_{bc}(\tit=0)=0$, consistent with the simulations, and the initial homogenisation of the control volume $\pl$ assumed in the model does not contribute to the infiltrated volume.} Hence, the effectiveness can be written as
\begin{equation}
    E(\tilde{t}) = 1 - \dfrac{3}{C_D}\dfrac{\Delta M_{\pl}/2+\Delta M_{\f 1} + \Delta M_{\f 2}}{(\rd-1)\tilde{t}}.
    \label{eq:effectiveness_analytical_2}
\end{equation}
Equation \eqref{eq:effectiveness_analytical_2} contains a key difference from the approach of \citet{bacot2022bubble} to define the infiltration flux from the saltwater to the freshwater sides. They assumed that the secondary gravity current starts from the location of the bubble curtain, while we consider that it emerges from the edge of the recirculation cell. As $\tit \to\infty$, this difference becomes negligible, but our definition is crucial for a correct estimation of $E$ at early and intermediate times.

In the spirit of the analytical model by \citet{bacot2022bubble}, a simplified expression for the steady-state limit at $\tilde{t}\to\infty$, \CR{that corresponds to the infinite domain asymptotic solution ($\tilde{L} \to \infty$)} can be obtained from Eq. \eqref{eq:effectiveness_analytical_2}, together with Eqs. \eqref{eq:DeltaML1} and \eqref{eq:DeltaMP}. This is possible because the semi-analytical model is formulated for infinite domains, since the domain length is not explicitly included in the model. However, for a finite domain, it can only be applied as long as the secondary gravity current has not reached the end wall of the tank. As $\tilde{t}\to\infty$ and considering that $\tilde{\bar{\rho}}_{\f 1}$ and $\tilde{\bar{\rho}}_{\pl}$ tend to a constant value in the steady-state limit, the expression Eq. \eqref{eq:effectiveness_analytical_2} simplifies to
\begin{equation}
    E(\tilde{t}\to\infty) \approx 1 -\dfrac{3 \Delta M_{\f 2}}{C_D (\rd -1) \tilde{t}}.
    \label{eq:approx_E}
\end{equation}
Hence, the contribution to the effectiveness of the recirculation cell becomes negligible with respect to that of the secondary gravity current that emerges from it. Furthermore, we use Eq. \eqref{eq:fluxes_fit_4a} while neglecting the initial transients in $\tilde{\bar{\rho}}_{\f 1}(\tilde{t})$ so that $\tilde{\bar{\rho}}_{\f 1}$ is considered constant equal to $\tilde{\bar{\rho}}_{\f 1}(\tilde{t}\to\infty)$. Then, the expression for $E(\tilde{t}\to\infty)$ can be further approximated as
\begin{equation}
    E(\tilde{t}\to\infty) \approx 1 - \dfrac{3 \alpha_{gc}}{C_D} \left[\frac{\tilde{\bar{\rho}}_{\f 1}(\tilde{t}\to\infty)-1}{\rd-1}\right]^{3/2}.
    \label{eq:approx_E_2}
\end{equation}

In the following sections, we present results from the semi-analytical model and compare it to those of the numerical simulations.  First, in Sect. \ref{section:res_density}, we focus on the density in the recirculation cells $\tilde{\bar{\rho}}_{\f 1}$ and $\tilde{\bar{\rho}}_{\s 1}$, and then on the effectiveness in Sect. \ref{section:res_effectiveness}.

\section{Results on the density changes inside the recirculation cells}
\label{section:res_density}

In this section, we present the solution of the system of equations Eqs. \eqref{eq:model_a}--\eqref{eq:model_c} for the densities in the control volumes $\f1$ and $\s 1$ to characterize the role of the parameters of the problem. Furthermore, we compare the results of the semi-analytical model with those of the numerical simulations to gain a better understanding of the latter.

 \begin{figure}
    \centering  \includegraphics[width = 1.0\textwidth]{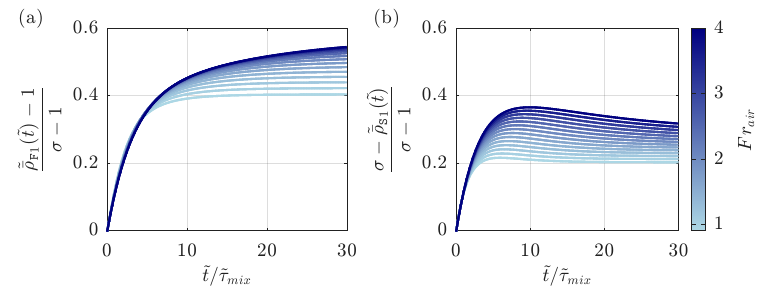}
 \caption{Time evolution of the normalised averaged density changes in the recirculation cells obtained from the semi-analytical model Eqs. \eqref{eq:model_a}--\eqref{eq:model_c}. The time has been normalised with the mixing time $\tilde{\tau}_{mix}=1/[0.15(\Fr+1)]$. The colour of the curves denote the value of $\Fr$.} 
      \label{fig:density_time_analytical}
\end{figure}

Figure \ref{fig:density_time_analytical} shows the values of the density changes in the control volumes $\f 1$ and $\s 1$, corresponding to the two recirculation cells, as a function of time obtained by solving Eqs. \eqref{eq:model_a}--\eqref{eq:model_c}. These density changes, given by $(\tilde{\bar{\rho}}_{\f 1} -1)/(\rd -1)$ and $(\rd-\tilde{\bar{\rho}}_{\s 1})/(\rd - 1)$, are normalised using the density difference and correspond in dimensional form to $(\bar{\rho}_{\f 1}-\rho_f)/\Delta \rho$ and $(\rho_s - \bar{\rho}_{\s 1})/\Delta \rho$. These two quantities are particularly important because the volume flow rates into control volumes $\f 2$ and $\s 2$ are proportional to them [see Eqs. \eqref{eq:fluxes_fit_3a} and \eqref{eq:fluxes_fit_3b}]. Furthermore, the normalised density \CR{change} in the control volume \CR{$\f 1$}, $(\tilde{\bar{\rho}}_{\f 1} -1)/(\rd -1)$, appears in Eq. \eqref{eq:effectiveness_analytical_2} to compute the effectiveness.

Each curve in Fig.~\ref{fig:density_time_analytical} corresponds to a different value of $\Fr$ (indicated by the colour) and therefore of the mixing time $\tilde{\tau}_{mix}=1/[0.15(\Fr+1)]$. Large values of $\tilde{\tau}_{mix}$ correspond to small values of $\Fr$. We observed that changing the value of $\rd$ does not affect the curves and therefore was here kept constant at $\rd=1.03$. This observation means that the differences in density within the recirculation cells and the original density ($\rho_f$ for the freshwater side or $\rho_s$ for the saltwater side) are proportional to $\Delta \rho$. 

Furthermore, the time in Fig.~\ref{fig:density_time_analytical} is normalised using the mixing time scale $\tilde{\tau}_{mix}$. For $\f1$, all the curves collapse into a single curve for early times $\tilde{t}/\tilde{\tau}_{mix}\lesssim 5$, while for $\s 1$, this occurs for $\tilde{t}/\tilde{\tau}_{mix}\lesssim 2.5$. The collapse of the different curves means that cases with lower values of $\tilde{\tau}_{mix}$ have a steeper ascent with time. This emphasises the physical interpretation of $\tilde{\tau}_{mix}$ where a smaller value indicates quick mixing and a larger value indicates slow mixing. Furthermore, since $\tilde{\tau}_{mix}=1/[0.15(\Fr+1)]$, cases with small $\Fr$ take longer to mix the density inside the recirculation cells. This is physically intuitive because a larger $\Fr$ value means a stronger bubble curtain, and hence enhanced mixing and faster homogenisation. For longer times, the curves no longer collapse and, as $\tilde{t}\to \infty$, each of them tends to a different constant value, which is higher for larger values of $\Fr$. These differences at longer times suggest that there is another physical process, with a different time scale, that dominates the evolution of density in recirculation cells after mixing inside the recirculation cells has taken place. This is the third time scale originally mentioned in the introduction, and, as will be made clear later, the related process is the transport of salt water by the secondary gravity currents.

 \begin{figure}
    \centering  \includegraphics[width = 1.0\textwidth]{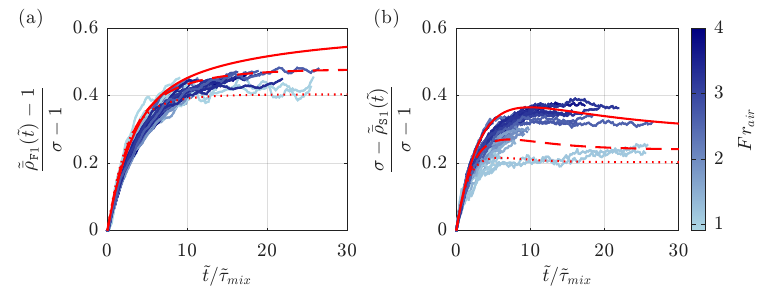}
 \caption{Normalised averaged density difference in the recirculation cells obtained from the numerical simulations (solid lines) as a function of time. The time has been normalised with the mixing time $\tilde{\tau}_{mix}=1/[0.15(\Fr+1)]$. The colour of the curves denote the value of $\Fr$. The smooth (red) curves correspond to solutions of the semi-analytical model Eqs. \eqref{eq:model_a}--\eqref{eq:model_c} for $\Fr=0.91$ (dotted), 1.5 (dashed) and 4 (solid) with $\rd=1.03$.}
\label{fig:density_time_numerical}
\end{figure}

Figure \ref{fig:density_time_numerical} shows the normalised density \CR{evolution} in the control volumes $\f 1$ and $\s 1$ obtained from the numerical simulations. To help compare with the results from the semi-analytical model, the solutions for $\Fr=0.91$, 1.5, and 4 ($\tilde{\tau}_{mix}\approx3.5$, 2.7, and 1.3) are also shown. The first clear observation is that most simulations stop before the density in the recirculation cells has reached a constant value even if some of the simulations use longer tanks ($L=4$ and 6 m) than in the experiments by \citet{bacot2022bubble} ($L=2$ m). This observation indicates that, particularly for large $\Fr$ values, reaching the steady state sometimes requires a longer time than is practically possible in simulations and experiments. A second observation is that the results from the simulations follow, in general, those of the semi-analytical model. The curves \CR{show reasonable} collapse for short times when normalising time with $\tilde{\tau}_{mix}$ and pull apart at longer times, with the simulations with smaller $\tilde{\tau}_{mix}$ (larger $\Fr$) values tending to larger density differences. However, the trends are not as clear in the numerical simulations because of fluctuations in the density that may be due to oscillations of the bubble curtain.

The presence of the two distinct behaviours in time warrants a closer examination using the semi-analytical model Eqs. \eqref{eq:model_a}--\eqref{eq:model_c}. Hence, in the following, we focus on two different limiting cases: short/initial times and the steady state ($\tilde{t} \to \infty$). Based on the flow evolution and the fact that mixing dominates for the short times, we assume that the secondary gravity currents have not yet formed or are negligibly weak. Instead, the main process is mixing within the control volumes $\f 1$, $\pl$ and $\s 1$. This means that the last terms on the right-hand side of Eqs. \eqref{eq:model_b} and \eqref{eq:model_c} are neglected. For the steady state, all terms on the right-hand side are included, but the time derivatives on the left-hand side of Eqs. \eqref{eq:model_a}--\eqref{eq:model_c} are set equal to zero.  

\subsection{Results for short/initial times}
\label{short_times_model}
For short times, we simplify Eqs. \eqref{eq:model_a}--\eqref{eq:model_c} assuming that the secondary gravity currents have not yet emerged or are weak enough so that $\tilde{q}_{\f 1 \to \f 2}= 0$ and $\tilde{q}_{\s 1 \to \s 2} = 0$ are a good approximation. As a result, the system of equations Eqs. \eqref{eq:model_a}--\eqref{eq:model_c} no longer depends on $\rd$ and only depends on $\Fr$. Still the initial conditions depend on $\rd$. After this simplification, the set of linear ordinary differential equations is solved analytically and the short-time ($st$) solution for $\tilde{\bar{\rho}}_{\f 1}$ is given by
\begin{equation}
\tilde{\bar{\rho}}^{st}_{\f 1}(\tilde{t}) = C_0 + C_{1,\f1} e^{-\tilde{t}/\tilde{\tau}_1} + C_{2,\f1}e^{-\tilde{t}/\tilde{\tau}_2},
\label{eq:model_short_times}
\end{equation}
where the time scales $\tilde{\tau}_1$ and $\tilde{\tau}_2$ depend only on $\Fr$ and the constants $C_0$, $C_{1,\f1}$, and $C_{2,\f1}$ depend on both $\Fr$ and $\rd$ (with $C_{2,\f1}<0$). The solutions for $\tilde{\bar{\rho}}_{\pl}$ and $\tilde{\bar{\rho}}_{\s 1}$ have the same form with $\tilde{\tau}_1$, $\tilde{\tau}_2$, and $C_0$ taking identical values for all control volumes, while $C_{1,i}$, and $C_{2,i}$ (with $i$ denoting the control volume) take a different value for each of them. The full expressions are given in Appendix \ref{section:appendix_short_times_solution}. For all control volumes, if secondary gravity currents do not arise, $\tilde{\bar{\rho}}_i\to C_0$ at $\tilde{t}\to \infty$, and there would be perfect mixing between the three cells. In fact, $C_0$ is equivalent to the density averaged over the volume at $\tilde{t}=0$, such that $C_0=(L_{\f 1}+L_\pl(1+\rd)/2+L_{\s 1}\rd)/(L_{\f 1}+L_\pl+L_{\s 1})$. However, this solution is strictly valid for short times, and the value of $\tilde{\bar{\rho}}_i$ deviates from $C_0$ at $\tilde{t}\to\infty$ due to the emergence of the secondary gravity currents.

\begin{figure}
    \centering
\includegraphics[width=0.98\textwidth]{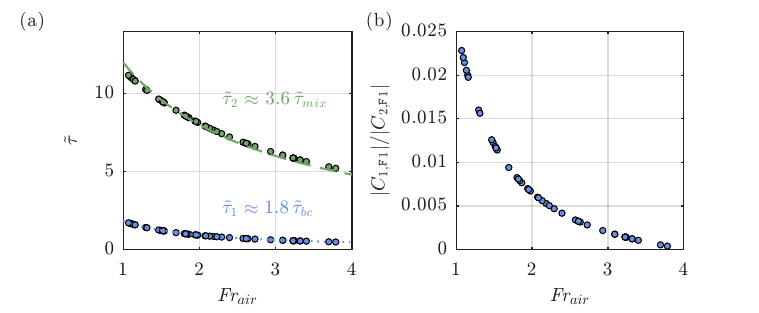}
      \caption{\CR{a) Variation of the e-folding times $\tilde{\tau}_1$ (blue bullets) and $\tilde{\tau}_2$ (green bullets) with $\Fr$, for the values of $\Fr$ used in the simulations. The fits $3.6\tilde{\tau}_{mix}$ (green dashed line) and $1.8\tilde{\tau}_{bc}$ (blue dotted line) are also shown.} b) Ratio of coefficients ($\left|C_{1,\f1}\right|/\left|C_{2,\f1}\right|$) plotted against $\Fr$} 
\label{fig:short_time_model}
\end{figure}

    The expression for $\tilde{\bar{\rho}}_{\f 1}^{st}$ in Eq. \eqref{eq:model_short_times} shows that the density grows as the sum of two exponential functions with $e$-folding time scales $\tilde{\tau}_1$ and $\tilde{\tau}_2$. Figure \ref{fig:short_time_model}(\emph{a}) shows the value of these time scales as a function of $\Fr$ used in the simulations. We observe that $\tilde{\tau}_2 \approx 3.6 \tilde{\tau}_{mix}$, where the proportionality factor was obtained using a linear fit. In addition, $\tilde{\tau}_2$ is an order of magnitude larger than $\tilde{\tau}_1$. Hence, we can conclude that $\tilde{\tau}_2$ is the time scale related to mixing in the recirculation cell. In contrast, $\tilde{\tau}_1$ being much smaller corresponds to the typical mixing time for the $\pl$ control volume around the bubble curtain. We define this time scale related to the bubble curtain as $\tau_{bc} = H/(gq_{air})^{1/3}$ (in dimensionless form $\tilde{\tau}_{bc} = \Fr^{-1}$), and from a linear fit, we obtain $\tilde{\tau}_1 \approx 1.8 \tilde{\tau}_{bc}$, which is also plotted in Fig.~\ref{fig:short_time_model}(\emph{a}) to further confirm that the smaller time scale is indeed the plume mixing time scale. To further assess the relative importance of the two terms with exponential functions in Eq. \eqref{eq:model_short_times}, we plot the ratio $\left|C_{1,\f 1}\right|/\left|C_{2,\f 1}\right|$ in Fig.~\ref{fig:short_time_model}(\emph{b}). Here, it can be seen that $|C_{1,\f 1}|\ll |C_{2,\f 1}|$. Therefore, the exponential function with the $e$-folding time $\tilde{\tau}_1$ has minimal influence on the solution and only at very short times, and it can be neglected, so that, to a good approximation,
\begin{equation}
\tilde{\bar{\rho}}^{st}_{\f 1}(\tilde{t}) \approx C_0 + C_{2,\f1}e^{-\tilde{t}/\tilde{\tau}_2}.
\label{eq:model_short_times_approx}
\end{equation}
This expression is in agreement with the scaling analysis in Sect.  \ref{section:scaling_analysis} in that the density in the recirculation cell is mixed with a typical time scale $\tilde{\tau}_2\propto \tilde{\tau}_{mix}$. Furthermore, this expression agrees with the collapse of the curves for short times observed in Figs.~ \ref{fig:density_time_analytical} and  \ref{fig:density_time_numerical} where time was normalised by $\tilde{\tau}_{mix}$.

\begin{figure}
    \centering
\includegraphics[width=0.98\textwidth]{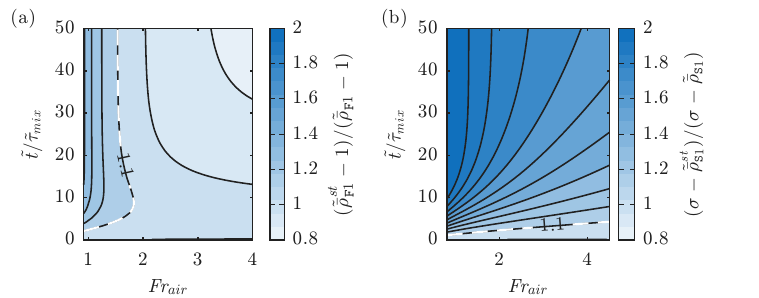}
      \caption{Ratio of the density difference in the recirculation cells from the short time approximation and the full semi-analytical solution as a function of $\Fr$ and $\tilde{t}/\tilde{\tau}_{mix}$ for a) the control volume $\f 1$ and b) the control volume $\s 1$. The dashed contours correspond to ratio values equal to $1.1$ as indicated for reference.} 
\label{fig:ratio_short_times}
\end{figure}

To determine more precisely the limit of the short-time approximation Eq. \eqref{eq:model_short_times}, in Fig.~\ref{fig:ratio_short_times}(\emph{a}), we plot the ratio  $(\tilde{\bar{\rho}}_{\f 1}^{st}-1)/(\tilde{\bar{\rho}}_{\f 1}-1)$ as a function of $\Fr$ and $\tilde{t}/\tilde{\tau}_{mix}$ with $\tilde{\bar{\rho}}_{\f 1}^{st}$ the short-time approximation and $\tilde{\bar{\rho}}_{\f 1}$ obtained from the full semi-analytical model by solving Eqs. \eqref{eq:model_a}--\eqref{eq:model_c}. Similarly, we plot the ratio $(\rd-\tilde{\bar{\rho}}_{\s 1}^{st})/(\rd-\tilde{\bar{\rho}}_{\s 1})$ in Fig.~\ref{fig:ratio_short_times}(\emph{b}). When these ratios are close to unity, the short-time ($st$) approximation can be considered applicable. As expected, for short times, these ratios are close to unity and, in general, the agreement holds for longer times as $\Fr$ grows. However, there is a clear difference between the accuracy of the short-time approximation in the control volumes $\f 1$ and $\s 1$. For $\f 1$, the short-term approximation holds relatively well (the corresponding ratio has a value of $1.0\pm0.1$) for all times shown for $2\lesssim \Fr \lesssim 4$. However, for $\s 1$, it is only valid for a short time with the ratio increasing above 1.1 for $\tilde{t}/\tilde{\tau}_{mix}$ greater than 1 to 5 depending on the value of $\Fr$. This difference between the two control volumes is also observed in Figs.~\ref{fig:density_time_analytical} and \ref{fig:density_time_numerical}, and it highlights the larger importance of the secondary gravity current on the saltwater side of the tank where the direction of this current is aligned with the surface current generated by the bubble curtain. However, to calculate the effectiveness of the bubble curtain, only the freshwater side is relevant, as established in Eq. \eqref{eq:effectiveness_analytical_2}.  

\subsection{Results for long times (steady state at $
\tilde{t}\to \infty $)}
Now we consider the second limiting case: the steady state as $\tilde{t}\to\infty$. In this case, the time derivatives on the left-hand side of Eqs. \eqref{eq:model_a}--\eqref{eq:model_c} are equal to zero, yielding a system of non-linear algebraic equations which is solved numerically. Furthermore, we derive a closed-form analytical solution similar to that obtained by \citet{bacot2022bubble} by assuming that the volume flow rates due to the secondary gravity currents  ($\tilde{q}_{\f 1 \to \f 2}$ and $\tilde{q}_{\s 1 \to \s 2}$) have become constant:
\begin{align}
|\tilde{q}_{\f 1 \to \f 2}| = |\tilde{q}_{\f2 \to \f 1}| = \gamma_{gc},\label{eq:fluxes_fit_5a}\\
|\tilde{q}_{\s1 \to \s2}| = |\tilde{q}_{\s2 \to \s 1}| = \kappa_{gc}.\label{eq:fluxes_fit_5b}
\end{align}
In this way, the non-linear terms in Eqs. \eqref{eq:model_a}--\eqref{eq:model_c} are linearised so that finding an analytical solution is possible. This assumption can be valid even if the density differences have barely reached a constant value (as seen in Fig.~\ref{fig:density_time_numerical})  because the flow rates are proportional to the square root of the density difference.

\begin{figure}
    \centering
\includegraphics[width=0.98\textwidth]{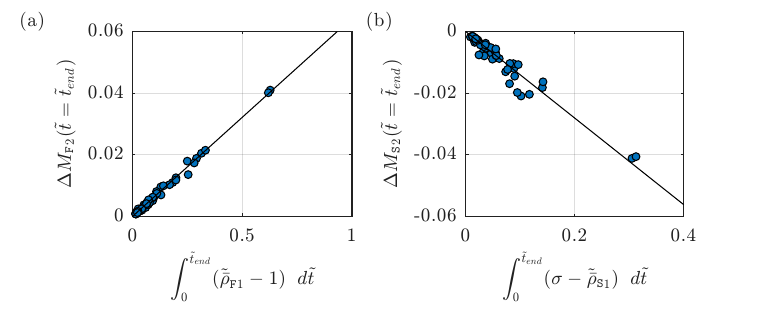}
      \caption{Change in salt mass in the control volumes $V_{\f 2}$ and $V_{\s 2}$ plotted against the respective theoretical estimates. a) Increase in salt mass in $V_{\f 2}$ and  b) decrease in salt mass in $V_{\s 2}$. The solid black lines represent the linear fits Eqs. \eqref{eq:fluxes_fit_6a} and \eqref{eq:fluxes_fit_6b}.} 
\label{fig:fluxes3}
\end{figure}

Following the same approach to estimate $\alpha_{gc}$ and $\beta_{gc}$ in Eqs. \eqref{eq:fluxes_fit_3a}--\eqref{eq:fluxes_fit_3b}, we consider the total amount of mass $M_{\f 2}$ and $M_{\s 2}$ gained or lost in control volumes $\f 2$ and $\s 2$ at the end of the simulations such that
\begin{align}
\Delta M_{\f 2}(\tilde{t} = \tit_{end})  =  \gamma_{gc}\int_{0}^{\tilde{t}_{end}} (\tilde{\bar{\rho}}_{\f 1}(\tilde{t}') - 1) d\tilde{t}',\label{eq:fluxes_fit_6a}\\
\Delta M_{\s 2}(\tilde{t} = \tit_{end})  =  -\kappa_{gc}\int_{0}^{\tilde{t}_{end}} (\rd-\tilde{\bar{\rho}}_{\s 1}(\tilde{t}') )d\tilde{t}'.
    \label{eq:fluxes_fit_6b}
\end{align}
The two sides of Eqs. \eqref{eq:fluxes_fit_6a} and \eqref{eq:fluxes_fit_6b} are plotted in Fig.~\ref{fig:fluxes3} and a linear relationship is observed, yielding fits for \CR{$\gamma_{gc}=0.064\pm0.001$} and \CR{$\kappa_{gc}= 0.14\pm0.008$}. The main differences between our model and the one proposed by \citet{bacot2022bubble} is that in their model, the volume flow rate due to the secondary gravity currents is modelled as a fraction of the flow entrained by the plume, such that $|\tilde{q}_{\f 1 \to \f 2}| = k |\tilde{q}_{\f 1 \to \pl}|$. The model solves for the unknown factor $k$. Furthermore, their model implicitly assumes a symmetry about the curtain or between the fresh and dense halves of the lock, which results in $|\tilde{q}_{\f 1 \to \f 2}| = |\tilde{q}_{\s 1 \to \s 2}|$ and $|\tilde{q}_{\f 1 \to \pl}| = |\tilde{q}_{\s 1 \to \pl}|$. However, the flow rate values obtained from the simulations show that this is not the case (see density field in Fig.~\ref{fig:analytical_model_sketch}). Hence, our model accounts for this asymmetry via the fit coefficients \CR{$\gamma_{gc}$} and $\kappa_{gc}$.

\begin{figure}
    \centering
\includegraphics[width=0.5\textwidth]{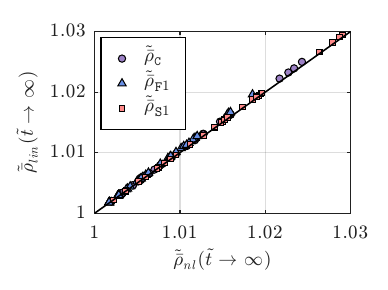}   \caption{Comparison of the densities in the control volumes $\f 1$, $\pl$, and $\s 1$ in the steady state as obtained using the numerical solution of the non-linear set of equations (represented by subscript $\textit{nl}$ in the abscissa) and analytical solution of the linearized set of equations (represented by subscript $\textit{lin}$ in the ordinate). The solid line represents the line $\tilde{\bar{\rho}}_{\textit{lin}}=\tilde{\bar{\rho}}_{\textit{nl}}$.} 
\label{fig:numerical_analytical_density_comparison}
\end{figure}

Using Eqs. \eqref{eq:fluxes_fit_1}\CR{--\eqref{eq:fluxes_fit_2} and \eqref{eq:fluxes_fit_5a}--\eqref{eq:fluxes_fit_5b}, with the values of $\gamma_{gc}$ and $\kappa_{gc}$ obtained from} \eqref{eq:fluxes_fit_6a}--\eqref{eq:fluxes_fit_6b}, the system of linear algebraic equations was solved analytically (see Appendix \ref{section:appendix_analytical_model} for details). To evaluate the accuracy of the linearisation, we compare, in Fig.~\ref{fig:numerical_analytical_density_comparison}, the solutions for the density in the control volumes $\f 1$, $\pl$, and $\s 1$ obtained by numerically solving the original non-linear set of equations and the analytical solution of the approximate linearised set of equations.  Excellent agreement is observed, showing that the linearisation is valid without compromising accuracy. 

\begin{figure}
    \centering
    \includegraphics[width=0.98\textwidth]{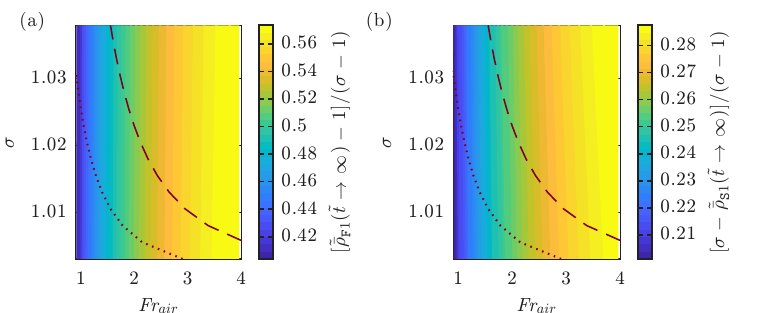}
   \caption{Densities $\tilde{\bar{\rho}}_{\f 1}$ and $\tilde{\bar{\rho}}_{\s 1}$ at steady state ($\tit \to \infty$) as a function of $\rd$ and $\Fr$. The red dashed line indicates the limit of $\rd$ and $\Fr$ values that can be achieved in the simulations \CR{(maximum air flow rate $q_{air}=0.005$ \unit{m^2.s^{-1}} with $H=0.15$ \unit{m})}, with values above the dashed line being inaccessible. \CR{The red dotted line corresponds to $\hat{q}_{air}=0.004$; values below it ($\hat{q}_{air}<0.004$) lie outside the range used to calibrate the semi-analytical model (see Sect.~\ref{section:flow_initialization}).}} 
\label{fig:s_Fr_map_rho_t_tilde_infinity}
\end{figure}

The normalised density differences in the recirculation cells $(\tilde{\bar{\rho}}_{\f1}-1)/(\rd -1)$ and $(\rd - \tilde{\bar{\rho}}_{\s1})/(\rd-1)$ in steady state are plotted in Fig.~\ref{fig:s_Fr_map_rho_t_tilde_infinity}. These quantities depend on $\rd$ only weakly, which means that, in dimensional terms, the density differences in the recirculation cells are, to a good approximation, proportional to the density difference $\Delta \rho$. Furthermore, we observe, as in Fig.~\ref{fig:density_time_analytical}, that the density differences are only a function of $\Fr$ with larger differences occurring for larger $\Fr$ values. 

\begin{figure}
    \centering  \includegraphics[width = 1.0\textwidth]{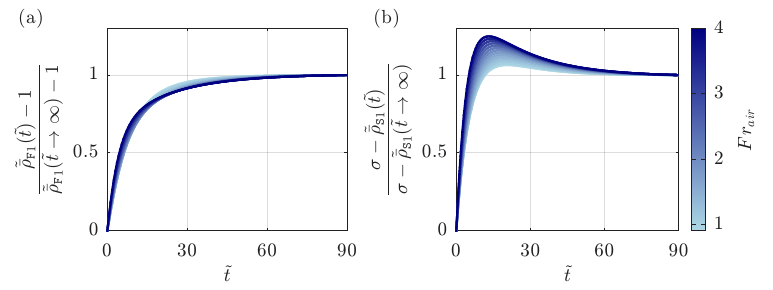}
 \caption{Average density difference in the recirculation cells normalised with the average density at $\tit\to \infty$ as obtained from the semi-analytical model Eqs. \eqref{eq:model_a}--\eqref{eq:model_c}. The colour denotes the value of $\Fr$.} 
      \label{fig:density_time_analytical_2}
\end{figure}

To determine the range of applicability, we plot the ratios of the density differences $(\tilde{\bar{\rho}}_{\f 1}(\tit)-1)/((\tilde{\bar{\rho}}_{\f 1}(\tit\to\infty)-1))$ and $(\rd-\tilde{\bar{\rho}}_{\s 1}(\tit))/(\rd-\tilde{\bar{\rho}}_{\s 1}(\tit\to\infty)$ as a function of the dimensionless time $\tilde{t}$ in Fig.~\ref{fig:density_time_analytical_2}. At early times, we do not observe a collapse of the curves, since the relevant time scale is $\tilde{\tau}_{mix}$. In general, the curves for these two ratios reach a value of one at different rates depending on $\Fr$. However, in general, this is observed to occur for all curves when $\tilde{t}\gtrsim 80$.

\section{Results on the effectiveness of bubble curtains}
\label{section:res_effectiveness}

\begin{figure}
    \centering
    \includegraphics[width=0.98\linewidth]{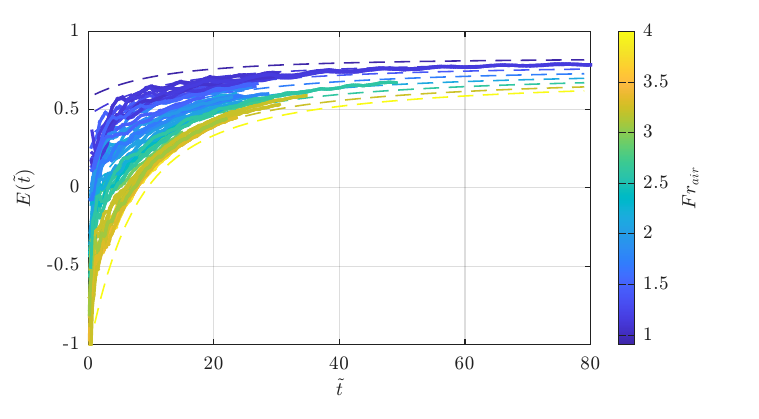}
    \caption{Effectiveness as a function of time. The solid lines correspond to the results from the numerical simulations, while the dashed lines in the background correspond to results from the semi-analytical model, see Eq. \eqref{eq:effectiveness_analytical_2}. The colour represents the value of $\Fr$.}
    \label{fig:E_vs_t}
\end{figure}

In this section, we present results for the effectiveness of bubble curtains using the semi-analytical model and numerical simulations. To understand these results, we build upon the insights into the density changes inside the recirculation cells presented in the previous section. Importantly, we address the reason for the scatter in the effectiveness values as a function of $\Fr$ observed in Fig.~\ref{fig:E_validation}.

Figure \ref{fig:E_vs_t} shows the effectiveness as a function of \CR{the dimensionless elapsed time after opening the lock gate (see Eq. \eqref{eq:t_tilde})} for all simulations and, in the background, for selected $\Fr$ values using the semi-analytical model. The calculations for the semi-analytical model were done with $\rd=1.03$, while for the simulations, several different values of $\rd$ were used as explained in Sect.\ref{section:flow_initialization}. In fact, it was discussed in the previous section that the relative changes in density in the control volume $\f 1$ and $\s 1$ given by $(\tilde{\bar{\rho}}_{\f1}-1)/(\rd -1)$ and $(\rd - \tilde{\bar{\rho}}_{\s1})/(\rd -1)$, respectively, are independent of $\rd$ to a good approximation. This result extends to the density change in the control volume $\pl$ given by $(\tilde{\bar{\rho}}_{\pl}-(\rd+1)/2)/(\rd -1)$ and, following Eq. \eqref{eq:fluxes_fit_4a}, for $\Delta M_{\f 2}/(\rd -1)$. Consequently, from Eqs. \eqref{eq:DeltaML1}--\eqref{eq:effectiveness_analytical_2}, it can be seen that the effectiveness as a function of time, $E(t)$, should also be, to a good approximation, independent of $\rd$. This result is remarkable because the quantity $\sigma -1$ [found in the denominator of the second term in Eq. \eqref{eq:effectiveness_analytical_2}] varies over one order of magnitude, from 0.003 to 0.04.

In general, both the simulations and the semi-analytical model show similar trends, and the colour correspondence between both sets of curves shows that there is also overall good quantitative agreement for similar values of $\Fr$. A more detailed quantitative comparison at early times ($\tilde{t}\lesssim 2$) is difficult to interpret because of differences between the simulations and the semi-analytical model. Firstly, in the semi-analytical model at $\tilde{t}$ = 0, it is assumed that the curtain region is well mixed and homogeneous $\tilde{\bar{\rho}}_{\pl}(\tilde{t} = 0) = (\rd + 1)/2$. However, in the simulations, on the freshwater side of the gate at $t=0$, $\tilde{\bar{\rho}}_{\pl}(\tilde{t}=0)=1$. Secondly, the semi-analytical model assumes that the densities in the control volumes $\f1$, $\pl$ and $\s 1$ are well mixed from $\tilde{t}=0$ but, in the simulations, this process takes some time (typically $\tilde{t}/\tilde{\tau}_{mix}\approx 1$). Finally, we observe that there are strong temporal fluctuations in the simulations that are not included in the semi-analytical model. We still note that, at early times, both simulations and the semi-analytical model with large $\Fr$ values result in negative $E$ values. These negative values are due to the enhanced transport of salt water toward the freshwater side due to the stronger mixing by the bubble curtain. \CR{The initialisation, whether in the simulations, the experiments or the semi-analytical model, only sets the distribution of a finite volume of salt water around the curtain at $\tilde{t}=0$, and its influence on $E$ fades as the flow evolves. A different initial distribution shifts the changes in mass $\Delta M_i$ in the numerator of Eq. \eqref{eq:effectiveness_analytical_2} by a constant, whereas the denominator grows linearly with $\tilde{t}$ because so does $V_o$ [Eq. \eqref{eq:v_open}]. The influence of the initial condition on $E$ therefore decays as $1/\tilde{t}$. In the semi-analytical model, for example, the assumed initial density in the control volume $\pl$ [Eq. \eqref{eq:DeltaMP}] changes $E$ by approximately $1/\tilde{t}$, i.e. by 0.1 at $\tilde{t}=10$. For the same reason, the simulations, in which the virtual “gate” is removed instantaneously, agree with the experiments, in which the gate is removed manually, after a short initial period (Appendix~\ref{section:appendix_validation}).} The general good agreement between the semi-analytical model and the simulations allows us to further study the time evolution of the effectiveness using the semi-analytical model, and a more detailed comparison for later times is given below.

\begin{figure}
    \centering
\includegraphics[width=0.98\textwidth]{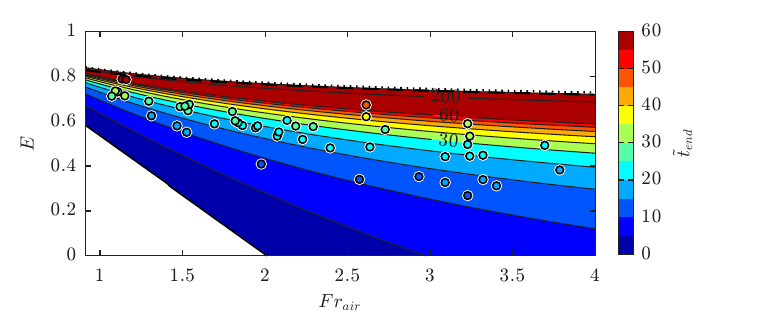}   \caption{Effectiveness $E$ vs $\Fr$ plot with the colour representing the end time $\tilde{t}_{end}$ at which the effectiveness is calculated. The simulation results are given by the bullets and the contour plot shows the results from the semi-analytical model. The analytical model for $E(\tilde{t}_{end} \to \infty)$ is given by the black dotted line. The dashed line shows the model for the steady state presented by \citet{bacot2022bubble}. To illustrate the typical time necessary to reach the steady state, the contour $\tilde{t}_{end}=200$ is presented.} 
\label{fig:E_vs_Fr_air_comparison_analytical}
\end{figure}

In the curtain-driven regime, larger values of $\Fr$ result in a lower effectiveness, as already expected from Fig.~\ref{fig:E_validation}, and it takes longer to reach the steady-state limit. Only a few simulations (mainly with $\Fr\approx 1$) can be considered to have reached this limit at the end time $\tilde{t}_{end}$ of the simulation. Hence, both time and $\Fr$ are crucial to determine the effectiveness of bubble curtains. To quantify this, we return to the plot of $E$ vs. $\Fr$ (Fig.~ \ref{fig:E_validation}) and add colour-coded information on the dependence of $E$ on the dimensionless end time $\tilde{t}_{end}$ \CR{(i.e. the time at which the secondary gravity current reaches the end wall of the tank, or the end time of the simulation if it does not, see Sect.~\ref{section:characteristics})} in Fig.~\ref{fig:E_vs_Fr_air_comparison_analytical}. As expected from Fig.~\ref{fig:E_vs_t}, markers denoting the results of the simulations are in good agreement with the results of the semi-analytical model presented as a colour map in the background. This result shows that the spread in the values of $E$ for a given $\Fr$ value as observed in the simulation results is explained by the dependence of $E$ on $\tilde{t}_{end}$.

Figure \ref{fig:E_vs_Fr_air_comparison_analytical} also shows the approximation for $E(\tilde{t}_{end}\to\infty)$ as given in Eq. \eqref{eq:approx_E_2} and the one proposed by \citet{bacot2022bubble}. Although our model considers additional complexity, such as an asymmetry in flow rates on the freshwater and saltwater sides of the lock, both approaches produce similar results for $E(\tilde{t}_{end}\to\infty)$ with only a minor offset. Hence, like for the experiments by \citet{bacot2022bubble}, the simulation results remain far from the limit value $E(\tilde{t}_{end}\to \infty)$, with the difference increasing with $\Fr$. Since our analysis holds only as long as none of the secondary gravity currents has reached an end wall, i.e. if $\tilde{t}_{end}\leq \tilde{t}_w$, the finite length of the domain translates into a finite time. It is then practical to translate the dependence of $E$ on $\tilde{t}_{end}$ to a dependence on $\tilde{L}$. In this way, the effectiveness is expressed as a function of the governing non-dimensional parameters of the problem: $\Fr$ and $\tilde{L}$. We do this by considering the time that the secondary gravity current on the freshwater side (and not the saltwater side) takes to reach the side wall because this is usually the relevant case in ship locks. However, a similar analysis can be performed for the saltwater side.

\begin{figure}
    \centering
\includegraphics[width=0.98\textwidth]{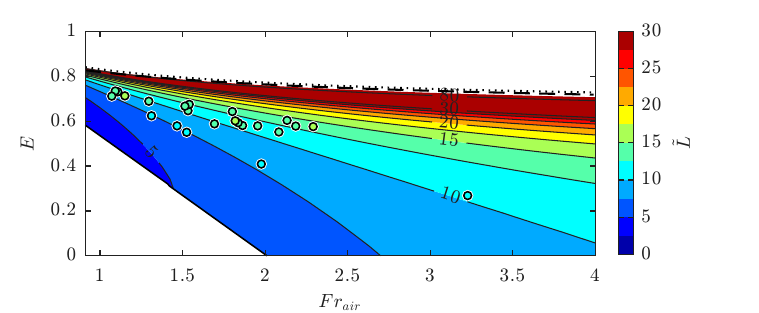}   \caption{ Effectiveness $E$ vs $\Fr$ plotted with the colour representing the dimensionless length of the domain $\tilde{L}$. The value of $E$ in this plot must be interpreted as the maximum possible effectiveness, provided that $\tilde{t}_{end}\leq \tilde{t}_w$. The simulation results are given by the bullets and the contour plot shows the results from the semi-analytical model. We only include simulations where the secondary gravity current has reached the end wall within the simulation time. The analytical model for $E(\tilde{t}_{end} \to \infty)$ is given by the black dotted line. The dashed line shows the model for the steady state presented by \citet{bacot2022bubble}.}  
\label{fig:E_vs_Fr_air_comparison_analytical_L}
\end{figure}

To link the finite length of the domain with the time $\tilde{t}_w$, we consider the propagation speed of the secondary gravity current:
\begin{equation}
\tilde{U}_{sgc}=2\alpha_{gc} \sqrt{\dfrac{\tilde{\bar{\rho}}_{\f 1}(\tit)-1}{\rd - 1}},
\end{equation}
following Eq. \eqref{eq:fluxes_fit_3a} and considering that the secondary gravity current has a height $H/2$. Then, the distance travelled by the secondary gravity current is
\begin{equation}
    \tilde{L}_{sgc}(\tilde{t})= 2\alpha_{gc}\int_0^{\tilde{t}} \sqrt{\dfrac{\tilde{\bar{\rho}}_{\f 1}(\tilde{t}')-1}{\rd - 1}}d\tilde{t}'
    \label{eq:twall}
\end{equation}
with $\tilde{t}'$ a dummy variable, and the time $\tilde{t}_w$ it takes for the gravity current to reach the wall can be obtained by solving the integral equation:
\begin{equation}
    \tilde{L}_{sgc}(\tilde{t}_w)=2\alpha_{gc}\int_0^{\tilde{t}_w} \sqrt{\dfrac{\tilde{\bar{\rho}}_{\f 1}(\tilde{t}')-1}{\rd - 1}}d\tilde{t}'=\tilde{L}/2 -\tilde{L}_{c, \f},
    \label{eq:twall2}
\end{equation}
where we have assumed that the gravity current starts to propagate at the end of the recirculation cell given by Eq. \eqref{eq:scaling_Lc_left}, that is, at $\tilde{x}=-\tilde{L}_{c,\f}$ at $\tilde{t}=0$. Through this equation, we can also relate the time $\tilde{t}_w$ to the length of the domain $\tilde{L}$, and obtain a figure similar to Fig.~\ref{fig:E_vs_Fr_air_comparison_analytical} but showing the dependence of $E$ on $\tilde{L}$ as shown in Fig.~\ref{fig:E_vs_Fr_air_comparison_analytical_L}. The value of $E$ in this graph must be interpreted as the maximum possible effectiveness, provided that $\tilde{t}_{end}\leq \tilde{t}_w$. Hence, this plot can also be useful for the design of future experiments, simulations, and ship locks. Considering still that $\tilde{t}_{end}\leq \tilde{t}_w$, by multiplying the corresponding value of $\tilde{L}$ by the dimensional depth of the system $H$, it is possible to determine the total length of the lock necessary to reach a certain value of $E$ given the value of $\Fr$. However, it is already clear from Fig.~\ref{fig:E_vs_Fr_air_comparison_analytical_L} that  $\tilde{L}=L/H\gtrsim 80$ (i.e. a very long domain) is necessary  for the value of $E$ to be close \CR{(within 0.02)} to the steady-state limit \CR{for $\Fr\gtrsim 2$, while $\tilde{L}\approx 25$ suffices close to the optimal value $\Fr\approx 0.91$}. Such values of $\tilde{L}$ are, in practice, difficult to achieve particularly for large $\Fr$ values in experiments, simulations, and ship locks. 

\section{Discussion}
\label{section:discussion}

\subsection{Towards a generalisation of the semi-analytical model}
In the current paper, we derived a semi-analytical model based on mass conservation using five control volumes: a central volume around the bubble curtain, two volumes encompassing the recirculation cells, and two volumes at the extremes where secondary gravity currents propagate as they emerge from the recirculation cells. The model reproduces well the temporal evolution of the key quantities, such as the average density within the recirculation cells and the effectiveness, as compared to those obtained from the numerical simulations. It is then useful as a fast surrogate model to study and understand the temporal evolution of the density field and the effectiveness of the bubble curtain.

The first process that takes place is the transport and homogenisation by the bubble curtain within the curtain itself and its immediate vicinity. This process has a typical time scale $\tau_{bc}\approx H/(g q_{air} )^{1/3}$. \CR{Within this time, the initial salinity field is spatially redistributed within the recirculation cells on both sides of the bubble curtain immediately after opening the lock gate}. Subsequently, the water in these recirculation cells is mixed, with a typical mixing time scale $\tau_{mix}\approx \sqrt{H/g'}/[0.15(\Fr+1)]$. Finally, secondary gravity currents emerge from the recirculation cells, with a typical time scale proportional to $\sqrt{H/g'}$, which is related to their propagation speed. During this later stage, salt water is slowly entrained into the recirculation cells and eventually seeps to the freshwater side. This final process is slow compared to the initial mixing. In practice, it is difficult to reach a steady state in domains of finite length because the secondary gravity currents reach the end walls before. However, the results of the semi-analytical model can be used to estimate the domain length and time needed for the flow to reach a steady state within a certain accuracy.

Even if the semi-analytical model proved useful in understanding the flow, it has limited predictive capabilities because the values of the model parameters ($|\tilde{q}_{\f1\to\pl}|$, $|\tilde{q}_{\s1\to\pl}|$, $\alpha_{gc}$, $\beta_{gc}$, $\tilde{L}_{c,\f}$, $\tilde{L}_{c,\s}$,) are determined from a specific configuration (e.g. bubble size, sparger design). However, a similar template could be used to construct a predictive mode if there is a better understanding of the relationship between parameter values and the governing processes. Still, from the value of the six empirical parameters, some are more robust than others as explained below.

First, the values of $|\tilde{q}_{\f1\to\pl}|$ and $|\tilde{q}_{\s1\to\pl}|$ given in Eqs. \eqref{eq:fluxes_fit_1} and \eqref{eq:fluxes_fit_2}, respectively, are in agreement with the results of \citet{bulson1961currents}, suggesting that their value might be independent of the bubble curtain configuration. However, it is still important to investigate whether they depend on the entrainment by the bubble curtain.

Second, for $\alpha_{gc}$ and $\beta_{gc}$, we are unaware of any experimental results, but we observe that $\alpha_{gc}< C_D/3$, where $C_D/3$ is the expected value for a gravity current directly arising from a lock exchange configuration (see Appendix \ref{section:Cd_open_cases}). This difference suggests that further research is needed to understand whether the flow inside the recirculation cell impacts the propagation of the secondary gravity currents.

 Finally, for the length of the recirculation cell $\tilde{L}_{c,\f}$ on the freshwater side of the lock, the observed trends as a function of $\Fr$ [i.e. $\tilde{L}_{c,\f}\propto\Fr/(\Fr+1)$] are also observed in the experimental results of \citet{bacot2022bubble}, but the actual size differs. This difference does not affect the good agreement between the models for the effectiveness at $\tit \to \infty$ because $\tilde{L}_{c,\f}$ is irrelevant in that limit. However, the effectiveness as a function of time should differ between our simulations and their experiments. Several reasons for the difference in the value of $\tilde{L}_{c,\f}$ have been discussed in Sect.  \ref{section:scaling_analysis} but, importantly, this difference also points to the need to better understand the relationship between the entrainment characteristics of the bubble curtain and the length of the recirculation cell. The entrainment by the bubble curtain can be modified by varying, for example, the bubble diameter and the sparger configuration \citep{fraga2016influence}. In fact, in a recent study, \citet{o2024effect} showed, using experiments for $\Fr$ values 0.8, 0.95 and 1.10, that the bubble size affects both the effectiveness and the optimal value of $\Fr$ for maximum effectiveness. This points to a dependence of the results obtained by \citet{bacot2022bubble} and the present study on certain characteristics of the bubble curtain, particularly the entrainment it induces. These aspects could further optimise the operation of bubble curtains in ship locks.

Conveniently, certain parameters are irrelevant for short or long times. In the short-time limit, $\alpha_{gc}$ and $\beta_{gc}$, which are related to flow rates due to the  secondary gravity currents, are irrelevant, while the lengths of the recirculation cells $\tilde{L}_{c,\f}$ and $\tilde{L}_{c,\s}$ are irrelevant in the steady state. Hence, if the interest lies in one of these limits, future investigations can focus on the relevant parameters.

\subsection{Implications for practical applications}

For practical applications in real ship locks, it is undesirable to operate at high $\Fr$ values (well into the curtain-driven regime) because the bubble curtain then requires more energy without further gain or even with a loss of effectiveness and the effectiveness can become negative. However, the results of the current paper are relevant for these applications because they describe and explain the flow with maximum effectiveness ($\Fr\approx 0.91$), which can be considered to be within the curtain-driven regime. Importantly, for this optimal value of $\Fr$, the effectiveness can change in the order of 20\% depending on the time that elapses after the opening of the gate. However, a similarly detailed description of the breakthrough regime is still warranted because it may be safer for ships sailing over it and it is more energy efficient. Hence, it could even be more advantageous than operating around maximum effectiveness.

It is difficult to quantify the accuracy of our simulations to represent the effectiveness in real ship locks under operational conditions because field measurements are difficult, and other physical processes (e.g., tides and the passage of ships) could influence the effectiveness. However, this study in a generic configuration helps to understand the basic principles of bubble curtains as a mitigation measure for salt intrusion. Still within the study of a generic and simplified configuration, there are several aspects that could affect the effectiveness of bubble curtains and deserve attention.

On the one hand, we neglected any variability in the bubbles that comprise the curtain by considering a constant and uniform bubble diameter, while in reality compressibility effects and polydispersity are the norm in real locks. However, we would expect these effects to only be relevant if they affect large scale entrainment properties of the curtain. On the other hand, we restricted our analysis to locks with an aspect ratio of $\tilde{W}\approx \CR{1.2}$. As representative of real ship locks, we take the locks in The Netherlands because they are numerous (more than 20 in total) and cover a broad range of dimensions, including the largest in the world located (the IJmuiden Sea Lock). For all these locks, $1.6\lesssim \tilde{W}\lesssim4$ when considering their depth with respect to the mean sea level. However, depth variations during the tidal cycle can further widen this range. For the wider locks, it can be expected from the work of \citet{riess1998recirculating} and \citet{raaghav7044620effect} that the flow becomes more three dimensional with horizontally-oriented recirculation cells along the width. Furthermore, the value of $\tilde{L}$ for the locks in The Netherlands ranges between 9 and 53, which \CR{largely overlaps} with the ranges covered in our numerical simulations (see section \ref{section:flow_initialization}) and highlights the importance of the semi-analytical model to estimate the effectiveness as a function of $\tilde{L}$.

\section{Conclusions}
\label{section:conclusions}
Using a combination of high-fidelity numerical simulations and semi-analytical modelling, we investigated the flow that develops in a lock-exchange configuration with a bubble curtain located at the gate position. The main goal was to understand and quantify the importance of the flow's time dependence on the effectiveness of the bubble curtain as a barrier for the intrusion of the denser salt water into the freshwater side.

The results of our numerical model agree well with the previous experimental results by \citet{bacot2022bubble}, showing a transition between the breakthrough and curtain-driven regimes occurring around the optimal value $\Fr\approx 0.91$ for maximum effectiveness. \CR{This study shows that the current numerical model can be used to gain valuable insights into the underlying physics of bubble curtains in lock-exchange configurations.} We focus on the curtain-driven regime, which is characterised by two recirculation cells (one on each side of the bubble curtain) and the emergence of secondary gravity currents from each of the cells. Importantly, the optimum value of $\Fr$, for which the highest effectiveness is achieved, can be considered to fall within this regime.

\CR{The novelty of this work lies in the description of the transient dynamics of the curtain-driven regime, which previous studies treated as a steady state. The simulations revealed the sequence of processes that govern the time evolution of the density field: the homogenisation by the bubble curtain, the mixing in the recirculation cells and the slow exchange through the secondary gravity currents. This understanding led to a scaling of the flow in the recirculation cells and to a semi-analytical model that extends the steady-state model of \citet{bacot2022bubble} to unsteady conditions and accounts for the asymmetry between the two sides of the curtain. The model provides a fast tool to estimate the effectiveness as a function of time and lock length, and shows that effectiveness values from experiments, simulations or field measurements can only be compared when they are reported together with the time or lock length to which they refer.}

Our results show that the effectiveness of bubble curtains in the curtain-driven regime depends strongly on both the air Froude number $\Fr$ and the time elapsed since opening the gate. The strong dependence on time explains the previously observed scatter when plotting effectiveness $E$ as a function $\Fr$. In conclusion, transient dynamics play a key role in determining the effectiveness of bubble curtains in the curtain-driven regime.

\section*{Acknowledgments} 
 The authors acknowledge financial support through the Perspective Program `SaltiSolutions' P18-32, in particular, Project 2 `Data and CFD for Solutions' (partly) financed by NWO Domain Applied and Engineering Sciences (2022/TTW/01344701). The simulations were run on the Dutch National Supercomputing Facility SURF through an NWO Domain Science grant (2024.006). The authors thank Dr. Daria Frank for kindly providing us with the experimental data for validation.

\section*{Author contributions} 
Shravan K. R. Raaghav: Conceptualization; Methodology; Investigation; Data Curation;  Formal analysis; Software; Visualization; Writing—original draft.\\\\
Ronald J. A. Driessen: Investigation; Visualization; Writing—review and editing.\\\\
Tom S. D. O’Mahoney: Conceptualization; Writing—review and editing.\\\\
Rob E. Uittenbogaard: Conceptualization; Writing—review and editing.\\\\
Herman J. H. Clercx: Conceptualization; Supervision; Project administration; Writing—review and editing; Funding acquisition.\\\\
Matias Duran-Matute: Conceptualization; Supervision; Investigation; Formal analysis; Writing—original draft; Writing—review and editing; Funding acquisition.

\section*{Data availability} 
The data will be made available by the authors upon reasonable request.

\section*{Declaration of interests}
The authors report no conflict of interest.

\appendix
\section{Flow rate for the open case}
\label{section:Cd_open_cases}
\begin{figure}
    \centering
\includegraphics[width=0.45\textwidth]{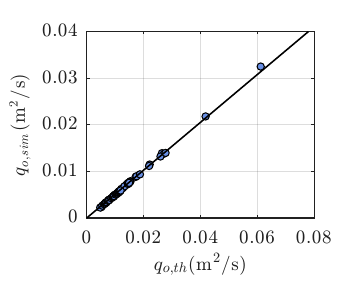}
    \caption{Theoretical infiltration flux $q_{o,th}$ for the open case plotted against the ones obtained from simulations $q_{o,sim}$. The solid black line is the linear fit.}  
    \label{fig:q_open_fit}
\end{figure}

For every combination of $\Delta \rho$ and $H$ chosen for the bubble curtain simulations, we additionally performed open (lock-exchange) simulations to compute the flow rate per unit width of dense water towards the freshwater side. We followed the same approach as
\citet{bacot2022bubble} and \citet{wilson1990gravity} to calculate the discharge coefficient ($C_D$) that accounts for the difference between the theoretical and actual flow rates. The theoretical flow rate per unit width is defined as $q_{o,th}=H\sqrt{g'H}/3$, such that $q_{o,th}=V_o/(W C_D t)$ with $V_o$ is defined in Eq. \eqref{eq:v_open}. Figure~\ref{fig:q_open_fit} shows the theoretical flow rate $q_{o,th}$ compared to the one obtained from the simulations $q_{o,sim}$. From a linear fit, we estimate $C_D = 0.513 \pm 0.004$. 

\section{Comparison of effectiveness values and density evolution profiles with laboratory experiments}
\label{section:appendix_validation}
 \begin{figure}
\includegraphics[width=0.9\textwidth]{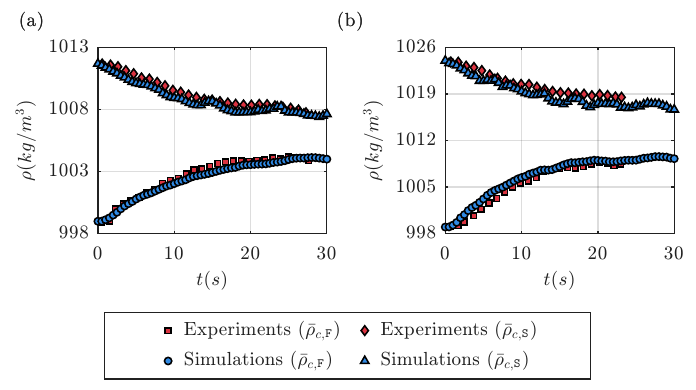}
    \caption{Temporal evolution of area-averaged density in the recirculation cells plotted for both the fresh ($\bar{\rho}_{c,\f}$) and dense ($\bar{\rho}_{c,\s}$) fluid half of the lock. The cases have the following input parameters: a) $Q_{air} = 5\times10^{-4}$ \unit{m^3.s^{-1}}, $\Delta \rho = 12.68$ \unit{kg.m^{-3}} and $H = 0.15$ \unit{m} ($\Fr=2.19$ and $\rd =1.012$); and b) $Q_{air}= 5\times10^{-4}$ \unit{m^3.s^{-1}}, $\Delta \rho = 25$ \unit{kg.m^{-3}} and $H = 0.15$ \unit{m} ($\Fr=1.51$ and $\rd =1.025$). For both cases, $W=0.2$ \unit{m} and $L=2$ \unit{m}. \CR{The values of $\Fr$ and $\rd$ are as reported by \citet{bacot2022bubble}.}} 
\label{fig:density_vs_time_validation}
\end{figure}
In Table \ref{table:effectiveness_comparison}, we show the values of effectiveness for select few cases where we exactly matched all the problem parameters (i.e. $L$, $W$, $H$, $\rho_s$, $\rho_f$ and $Q_{air}$) so that all the non-dimensional groups match that of \citet{bacot2022bubble}. As seen in Table \ref{table:effectiveness_comparison}, we observe excellent agreement with the experimental data. 

\begin{table}[t]
\centering
\caption{Effectiveness values $(E(t = t_{end}))$ comparison between simulations ($E_{sim}$) and experiments ($E_{exp}$) for cases matching the input parameters of \citet{bacot2022bubble}.}
\label{table:effectiveness_comparison}
\begin{tabular*}{\textwidth}{@{\extracolsep{\fill}}cccccc}
\hline
$\Fr$ & $\sigma$ & $\tilde{L}$ & $\tilde{W}$ & $E_{sim}$ & $E_{exp}$\\
\hline
1.09 & 1.0367 & 13.33 & 1.33 & 0.734 & 0.748\\
1.29 & 1.0342 & 13.33 & 1.33 & 0.69 & 0.713\\
1.48 & 1.0125 & 13.33 & 1.33 & 0.66 & 0.66\\
1.54 & 1.0384 & 13.33 & 1.33 & 0.675 & 0.677\\
\hline
\end{tabular*}
\end{table}

In Fig.~\ref{fig:density_vs_time_validation}, we plot the temporal evolution of the density averaged within the recirculation cells on both sides of the lock and compare it with the experimental data of \citet{bacot2022bubble}. We considered two cases with all the same input parameters (i.e. $q_{air}$, $\Delta \rho$, $L$, $W$, and $H$) as the ones used in the experiments. The RMS error for $\bar{\rho}_{c,\f}$ and $\bar{\rho}_{c,\s}$ for both cases shown in Fig.~\ref{fig:density_vs_time_validation} is \CR{$\lesssim 1$ \unit{kg.m^{-3}}}. In the experiments, the density measurements have an uncertainty of 1 \unit{kg.m^{-3}}. Hence, the agreement can be considered as excellent.

\section{$L_c$ computation}
\label{section:appendix_L_cell}
To obtain the value of $L_c$ from the simulations, we followed the method proposed by \citet{bacot2022bubble} with minor adjustments. First, an estimate of $L_{c,\f}$ is made based on visual inspection of animations of the width-averaged density fields by finding the location where the density field at the outer region of the recirculation cell matches $\rho_s$. For this step, a rough estimate is sufficient since it has a negligible influence on the final value of $L_{c,\f}$. This first approximation $L_{c,\textit{est}}$ is used to estimate the average density in the upper half of the recirculation cell in the freshwater side of the lock, i.e., for $-L_{c,\textit{est}}\leq x\leq0$ and $H/2\leq y \leq H$.  For every $y$ position within this range, the $x$ position where $\rho\leq (\bar{\rho}_{c, \f} + \rho_f)/2$ is identified. The identified $x$ positions are averaged resulting in a height-averaged recirculation cell length. This process is repeated from the time the secondary gravity current is first observed to the time it reaches the wall. The resulting $L_{c,\f}$ values are averaged, and the mean value is the final measure of $L_{c,\f}$ presented in Fig.~\ref{fig:L_cell_non_dimensional}. The difference between the present method and that of \citet{bacot2022bubble} is that they randomly chose ten frames in the steady phase of the curtain-driven regime and computed $L_{c,\f}$ for the averaged flow field. The uncertainty obtained by \citet{bacot2022bubble} was the uncertainty in the height-averaged $L_{c,\f}$ due to the spatial variation in $L_{c,\f}$ with height. In contrast, in our case, the uncertainty originates from the temporal variation of $L_{c,\f}$. To obtain the length of recirculation cell in the saltwater side of the lock, a similar procedure was followed, but instead of considering the upper half of the recirculation cell, we considered the lower half. 


\section{Procedure to compute flow rates}
\label{section:appendix_flux_compute}
\begin{figure}
    \centering
\includegraphics[width=0.8\textwidth]{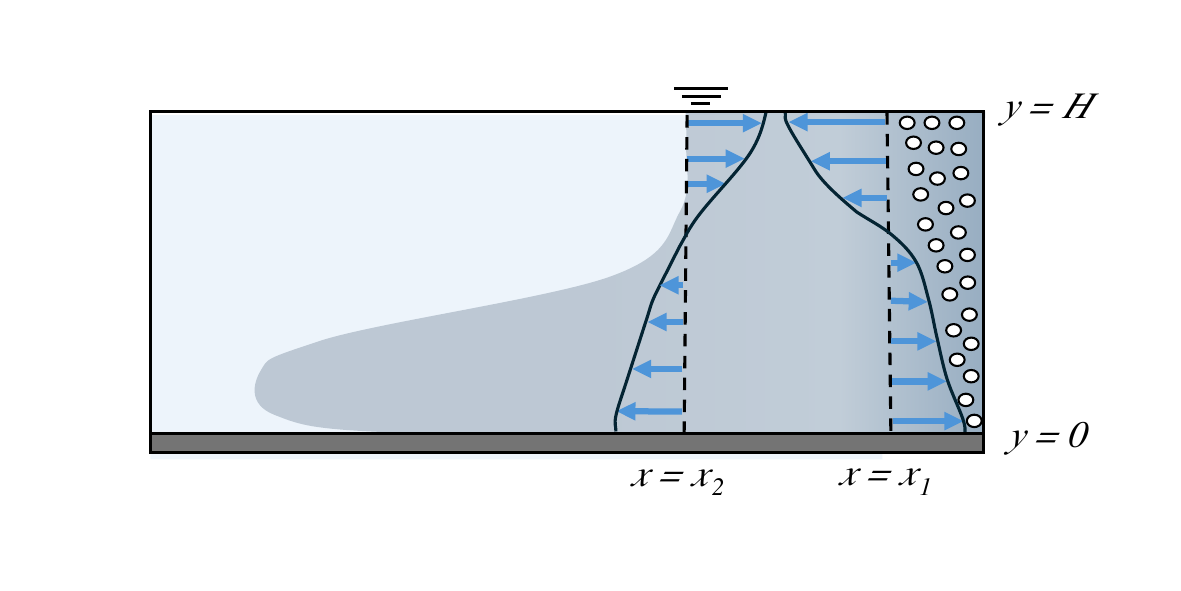}
    \caption{Sketch of the curtain and the recirculation in the freshwater side of the lock showing the vertical lines sliced to compute the fluxes. Here $x_1 = - 0.35H$ and $x_2 = -L_{c,\mathcal{\f}}$.}
\label{fig:flux_compute}
\end{figure}

The flow rates used to estimate the coefficients for the semi-analytical model were computed from the time-averaged and width-averaged velocity fields from the simulations. From these velocity fields, velocity profiles at $x = - 0.35H$, $x = 0.35H$, $x = - L_{c,\mathcal{\f}}$, $x =L_{c,\mathcal{\s}}$ are extracted to compute the flow rates $q_{\f1 \to \pl}$, $q_{\s1 \to \pl}$, $q_{\f1 \to \f2}$ and $q_{\s1 \to \s2}$, respectively. A typical schematic of the bubble curtain and the velocity profiles on the freshwater side of the curtain is shown in Fig.~\ref{fig:flux_compute}.  
The flow rate per unit width is computed by integrating the velocity field in the vertical from $y = 0$ to $y = H$ so that, considering volume conservation,
\begin{equation}
|q_{i\to j}| = \dfrac{1}{2}\int_{0}^{H} {|v_x|}dy.
\label{eq:flux_compute}
\end{equation}

\section{Analytical solutions for short times and in the steady state limit}
\label{section:appendix_short_times_solution}
\subsection{Analytical solution for short times}
Here, we provide the full-form solutions in the short-times approximation for the average densities: $\tilde{\bar{\rho}}_{\pl}(\tilde{t})$, $\tilde{\bar{\rho}}_{\f 1}(\tilde{t})$ and $\tilde{\bar{\rho}}_{\s 1}(\tilde{t})$. The form of these solutions is given in Eq. \eqref{eq:model_short_times}. The $e$-folding times, $\ti{\tau}_1$ and $\ti{\tau}_2$ are the same for the three densities and are given by
\begin{align}
\ti{\tau}_1 = \frac{2}{A+B+C+D+K},
\end{align}
and
\begin{align}
\ti{\tau}_2 = \frac{2}{A+B+C+D-K},
\end{align}
where $A=|\tilde{q}_{\f 1 \to \pl}|/\tilde{L}_{\pl}$, $B=|\tilde{q}_{\s 1 \to \pl}|/\tilde{L}_{\pl}$, $C=|\tilde{q}_{\f 1 \to \pl}|/\tilde{L}_{\f 1}$, $D=|\tilde{q}_{\s 1 \to \pl}|/\tilde{L}_{\s 1}$, $S = BC + (A+C)D$ and $K = \sqrt{A^2 + 2 A (B + C - D) + (B - C + D)^2}$. 
The constant $C_0$ is also the same for the three densities and is given by
\begin{equation}
C_0 = \frac{ 
   2 B C \rd + 2 A D + C D \left(\rd + 1\right) }{2S}.
\end{equation}
The other two coefficients multiplying the exponential functions are different for each control volume, and they are given by

\begin{equation}
    C_{1,\pl} =  \frac{1}{{4 SK}}[A D (K+A) + A \left(C - D\right) \left(3 B + D\right)  
    - 
   B C \left(B - C + D +  K\right)]\left(\rd - 1\right),
\end{equation}

\begin{equation}
    C_{2,\pl} =  \frac{1}{{4 SK}}[A D (K-A) - A \left(C - D\right) \left(3 B + D\right) + 
   B C \left(B - C + D -  K\right)]\left(\rd - 1\right),
\end{equation}



\begin{equation}
    C_{1,\f1}  =  \frac{C}{4SK} \left[2B^2 -D(  A - 3B + C - D+K) +2B(A- K) \right]  \left(\rd-1\right),
\end{equation}

\begin{equation}
    C_{2,\f1}  =  - \frac{C}{4SK} \left[2B^2 -D( A - 3B + C - D-K) +2B(A + K) \right]  \left(\rd-1\right),
\end{equation}


\begin{equation}
    C_{1,\s1} = -\frac{D}{4SK} \left[2 A^2 + 
   C \left(3A-B + C - D - K\right) + 
   2A \left(B - K\right)\right]  \left(\rd-1\right),
\end{equation}

\begin{equation}
    C_{2,\s1} = \frac{D}{4SK} \left[2 A^2 + 
   C \left(3A-B + C - D + K\right) + 
   2A \left( B + K\right)\right]  \left(\rd-1\right),
\end{equation}

\subsection{Analytical solution at steady state conditions}
\label{section:appendix_analytical_model}
In the steady state scenario for the semi-analytical model Eqs. \eqref{eq:model_a}--\eqref{eq:model_c}, the time derivatives vanish, which yields 
\begin{align}\label{eq:model_steady}
 -\tilde{q}_{\f 1 \to \pl}(\tilde{\bar{\rho}}_{\pl}-\tilde{\bar{\rho}}_{\f 1})- \tilde{q}_{\s 1 \to \pl}(\tilde{\bar{\rho}}_{\pl}-\tilde{\bar{\rho}}_{\s 1}) = 0, \\
\tilde{q}_{\f 1 \to \pl}(\tilde{\bar{\rho}}_{\pl}-\tilde{\bar{\rho}}_{\f 1})+ \tilde{q}_{\f 1 \to \f 2}(1-\tilde{\bar{\rho}}_{\f 1}) = 0,\\
 \tilde{q}_{\s 1 \to \pl}(\tilde{\bar{\rho}}_\pl-\tilde{\bar{\rho}}_{\s 1})+ \tilde{q}_{\s 1 \to \s 2}( \rd-\tilde{\bar{\rho}}_{\s 1}) = 0,
\label{eq:mass_conservation_steady_state}
\end{align}
where, for brevity, $\tilde{q}$ denotes the absolute value of the volume flow rates.

The resulting system of equations is presented in matrix form as
\begin{gather}
\begin{pmatrix}
-\tilde{q}_{\f1 \to \pl}-\tilde{q}_{\s1 \to \pl} & \tilde{q}_{\f1 \to \pl} & \tilde{q}_{\s1 \to \pl}\\
\tilde{q}_{\f1 \to \pl} & -\tilde{q}_{\f1 \to \pl}-\tilde{q}_{\f1 \to \f2} & 0\\
\tilde{q}_{\s1 \to \pl} & 0 & -\tilde{q}_{\s1 \to \pl}-\tilde{q}_{\s1 \to \s2}
\end{pmatrix}
\cdot
\begin{pmatrix}
\tilde{\bar{\rho}}_{\pl}\\
\tilde{\bar{\rho}}_{\f 1}\\
\tilde{\bar{\rho}}_{\s 1}
\end{pmatrix}=
\begin{pmatrix}
0\\
-\tilde{q}_{\f1 \to \f2}\\
-\rd \tilde{q}_{\s1 \to \s2}
\label{eq:mass_conservation_matrix}
\end{pmatrix}.
\end{gather}

We substitute for the volume flow rates from Eqs. \eqref{eq:fluxes_fit_1}--\eqref{eq:fluxes_fit_2} and Eqs. \CR{\eqref{eq:fluxes_fit_5a}--\eqref{eq:fluxes_fit_5b}} into Eq. \eqref{eq:mass_conservation_matrix} and solve the system of equations yielding





\begin{equation}
\begin{aligned}
\tilde{\bar{\rho}}_{\pl} = \frac{\CR{\gamma_{gc}} M (\CR{\kappa_{gc}} + N) + 
   \CR{\kappa_{gc}} (\CR{\gamma_{gc}} + M) N \rd}{\CR{\gamma_{gc}} M N + 
   \CR{\kappa_{gc}} M N + 
   \CR{\gamma_{gc}} \CR{\kappa_{gc}} (M+N)},
\end{aligned}
\end{equation}

\begin{equation}
\begin{aligned}
\tilde{\bar{\rho}}_{\f 1} = \frac{\CR{\gamma_{gc}}M N + 
   \CR{\gamma_{gc}} \CR{\kappa_{gc}} (M+N) + 
   \CR{\kappa_{gc}} M N \rd}{\CR{\gamma_{gc}} M N + 
   \CR{\kappa_{gc}} M N + 
   \CR{\gamma_{gc}} \CR{\kappa_{gc}} (M+N)},
\end{aligned}
\end{equation}

\begin{equation}
\begin{aligned}
\tilde{\bar{\rho}}_{\s 1} = \frac{\CR{\gamma_{gc}} M N + 
   \CR{\kappa_{gc}} M N \rd + 
   \CR{\gamma_{gc}} \CR{\kappa_{gc}} (M+N) \rd}{\CR{\gamma_{gc}} M N + 
   \CR{\kappa_{gc}} M N + 
   \CR{\gamma_{gc}} \CR{\kappa_{gc}} (M+N)},
\end{aligned}
\end{equation}
where
\begin{equation*}
M = (-0.06 + 0.19 \Fr),
\end{equation*}
\begin{equation*}
N = (0.02+ 0.17 \Fr).
\end{equation*}
\bibliography{bubble_curtain_dynamics_1}

\end{document}